\documentclass[aip,prl,reprint,showkeys,floatfix, graphics,groupedaddress,superscriptaddress]{revtex4-2} 
\usepackage[margin=0.8in]{geometry}
\newcommand{\comment}[1]{}
\usepackage[english,french]{babel}
\usepackage{printlen}
\usepackage{color}
	\usepackage[toc,page]{appendix}
	\usepackage[T1]{fontenc}
	\usepackage[colorlinks=true,allcolors=blue]{hyperref}
	\usepackage{url}
	\usepackage[pdftex]{graphicx}
	\usepackage{amsmath}
	\usepackage{amssymb}
	\usepackage{appendix}
	\usepackage[justification=justified]{caption}
        \usepackage{ragged2e}
	\usepackage{subcaption}
	\usepackage{float}
        \usepackage{placeins}
	\usepackage{amsmath}
	\usepackage{enumitem}
	\usepackage{makecell}
	\usepackage{txfonts}
	\usepackage{setspace}
	\usepackage[svgnames]{xcolor}
	\usepackage{hhline}
	\usepackage{tabu}
	\usepackage[multidot]{grffile}
	\definecolor{klein}{RGB}{0,47,167}
	\definecolor{malachite}{RGB}{31,160,85}
	\definecolor{DBlue}{rgb}{0.0, 0.0, 1.0}
	\definecolor{DGreen}{rgb}{0.5, 1.0, 0.0}

	\renewcommand{\thefootnote}{\Roman{footnote}} 

	\newcommand{\Fref}[1]{Figure~\ref{#1}}
	
	\newcommand{\fref}[1]{figure~\ref{#1}}

	\newcommand{\Eref}[1]{Equation~(\ref{#1})}
	
	\newcommand{\eref}[1]{equation~(\ref{#1})}

	\newcommand{\Cref}[1]{Chapter~(\ref{#1})}
	
	\newcommand{\cref}[1]{chapter~(\ref{#1})}

	\newcommand{\cross}[2]{{\boldsymbol{#1}}\times{\boldsymbol{#2}}} 

	\newcommand{\RLT}[1]{-R_0\nabla T_{#1}/T_{#1}}
	\newcommand{\RLN}[1]{-R_0\nabla n_{#1}/n_{#1}}

\begin{document}
\selectlanguage{english}
\title{Is Turbulence able to Generate Magnetic Islands in Tokamaks? Gyrokinetic Simulations of Turbulence-Driven Magnetic Islands in Toroidal Geometry}
\author{F.~Widmer}
\affiliation{Max Planck Institute for Plasma Physics, 85748 Garching, Germany}
\email{fabien.widmer@ipp.mpg.de}
\author{E.~Poli}
\affiliation{Max Planck Institute for Plasma Physics, 85748 Garching, Germany}
\author{A.~Mishchenko}
\affiliation{Max Planck Institute for Plasma Physics, 17491 Greifswald, Germany}
\author{A.~Ishizawa}
\affiliation{Graduate School of Energy Science, Kyoto University, Uji, Kyoto 611-0011, Japan}
\author{T.~Hayward-Schneider}
\author{A.~Bottino}
\affiliation{Max Planck Institute for Plasma Physics, 85748 Garching, Germany}
\date{\today}
\begin{abstract}
We report a universal mechanism for turbulence-driven magnetic islands in fusion plasmas. Using gyrokinetic simulations of a linearly stable tearing mode
in a large-aspect-ratio toroidal geometry under collisionless conditions, we demonstrate that micro-instabilities generate an
$\cross{E}{B}$ flow that drives magnetic field line reconnection. This process forms multiple small-scale islands along the resonant
surface, which interact nonlinearly and eventually coalesce into large-scale magnetic islands. These islands are capable of significantly flattening the
equilibrium profile across the island O-points and thus acting as a seed for their further neoclassical growth.
Notably, the mechanism operates independently of the parity of the destabilizing micro-instability. 
\end{abstract}

\maketitle
\begin{figure}
	\begin{center}
	\includegraphics[width=\linewidth,keepaspectratio]{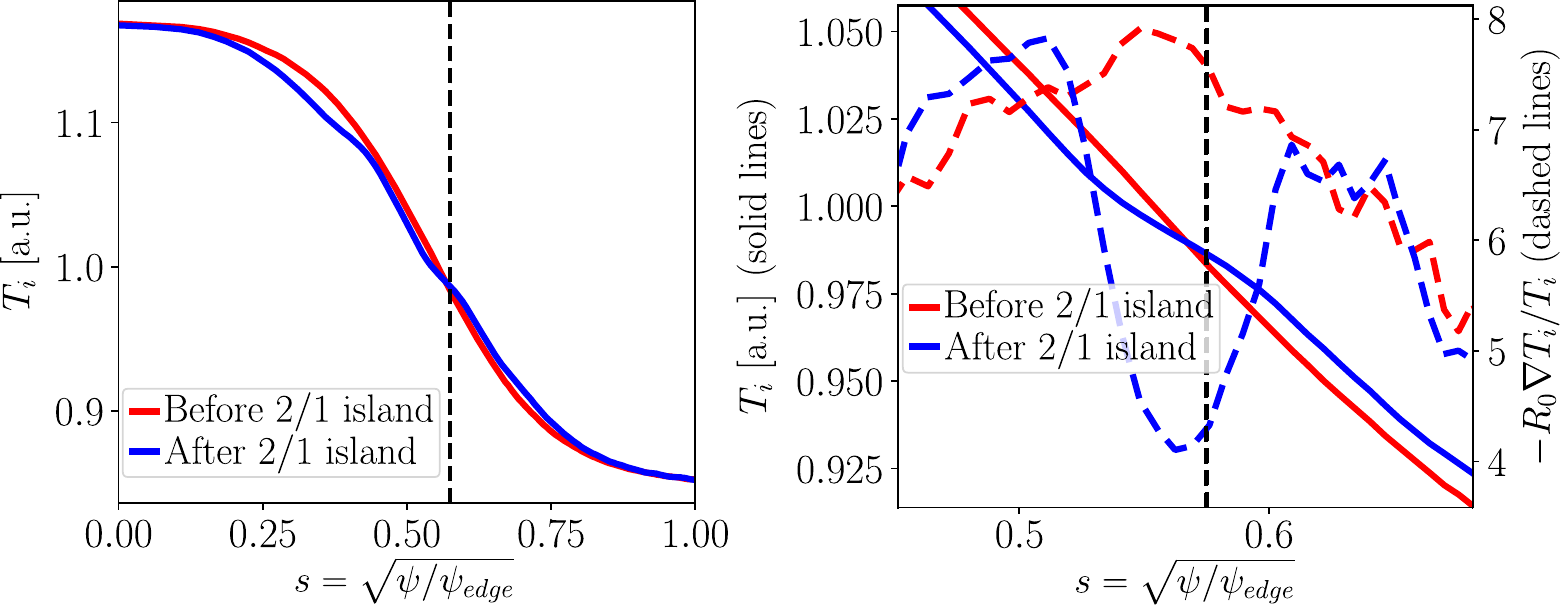}
		\caption{\justifying ITG case ($\RLN{}=\RLT{e}=0$, $\RLT{i}=8$, $\beta=0.05\%$): flux surface averaged temperature profiles (left)
                         and normalized gradients scale length (right). After the $2/1$ island emergence, the profiles flatten around
                         the rational surface (vertical black dashed-line).}
        \label{fig:MTMTeProfs}
	\end{center}
\end{figure}

Optimal performance in fusion reactors is achieved for high plasma beta ($\beta = \mu_0 n T / B^2$), the ratio of plasma to magnetic pressure.
However, this comes with a significant challenge: controlling magnetohydrodynamics (MHD) scale instabilities, particularly the growth of Neoclassical Tearing Modes (NTMs). Non-linearly, NTMs form radially localized magnetic islands that can lead to confinement deterioration, rapid plasma termination and potentially damage the reactor vessel.\cite{IdaPRL18,BardocziPRL21,BardocziNF24}
Many aspects of the nonlinear evolution of NTMs,\cite{kotschenreuther_PF1985,Carreras1981a,SauterPoP97a,SweeneyNF17}
and their control,\cite{ZohmNF99,PraterNF07,Isayama_NF2009} are well understood. However, the characterization of the mechanisms that govern magnetic
islands in their early phase, and their seeding, remains a challenge.\cite{IsayamaJPFR13,PoliPPR16,IshizawaPPCF19,Choi2021a}
While magnetic islands can grow from linearly unstable tearing modes ($\Delta' > 0$),\cite{furth_PF1973} experimental observations
show that islands often form under $\Delta' < 0$ conditions.\cite{kotschenreuther_PF1985,Carreras1981a,SauterPoP97a,SweeneyNF17}
 In such cases, some triggering event is usually needed to provide a seed island that is wide
enough to flatten the pressure profile and activate the neoclassical drive.\cite{SauterPoP97a}
On the other hand, seed islands may arise from nonlinear coupling between turbulence and magnetic reconnection,\cite{MuragliaPRL11,IshizawaJPP15}
which could offer an explanation for the appearance of ``triggerless'' islands in some experiments.\cite{FietzPPCF13}
\newline\indent In tokamaks, a suitable framework for investigating the mutual interaction between magnetic islands and turbulence is provided by the gyrokinetic
theory.\cite{BrizardRMP07} Previous studies of imposed magnetic islands have shown that radially large islands can suppress turbulence transport around
them due to the combination of zonal and island-induced flows.\cite{PoliNF09,PoliPPCF10a,JiangPoP14,ZarzosoNF15,MutoPoP22,LiNF23}
In the case of the self-consistent growth of small and large-scale magnetic perturbations in gyrokinetic simulations,
investigations focused on the impact of turbulence on linearly unstable tearing modes ($\Delta' > 0$).\cite{HornsbyPPCF15,HornsbyPPCF16,JitsukNF24,WidmerPoP24,WeiNF25} In this paper, we focus for the first time on the characterization of "turbulence-driven" magnetic islands in tokamak geometry. In the simplified
setup detailed below, different micro-instabilities are excited by varying the density $n$, electron temperature $T_e$, ion temperature $T_i$ gradients and the plasma $\beta$.
A consistent picture emerges showing how large-scale magnetic islands eventually grow as a result of the coalescence of small-scale islands.
These can produce a noticeable flattening of the pressure profile, acting as a seed for their further (neoclassical) growth.
\begin{figure}[htp]
	\begin{center}
	\includegraphics[width=0.6\linewidth,keepaspectratio]{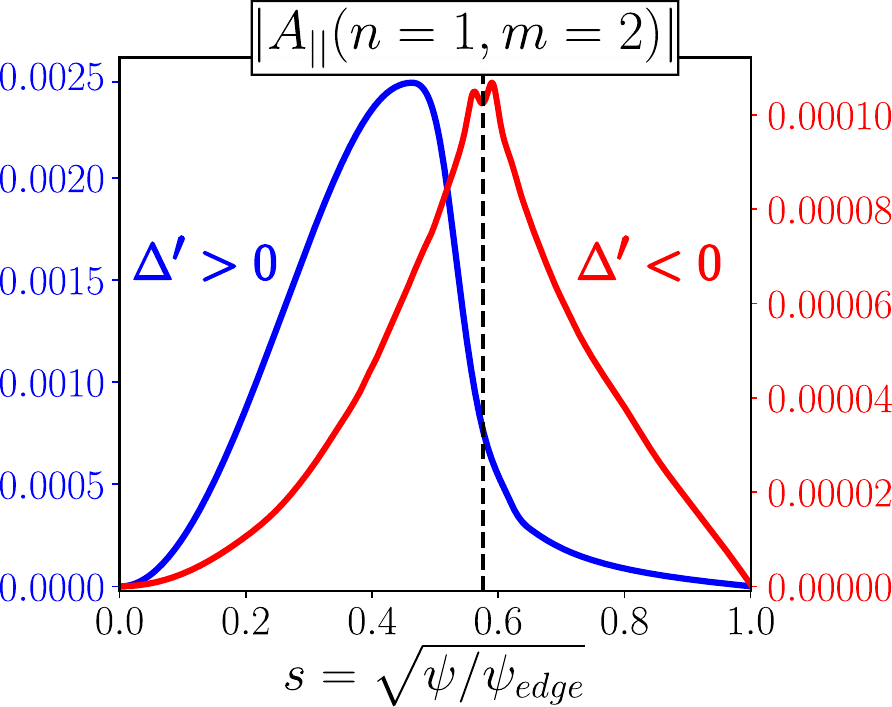}
        \caption{\justifying $|A_{||}(m/n=2/1)|$ eigenfunction in toroidal geometry for a current-driven islands
                 ($\Delta'>0$) in blue and a turbulence-driven one ($\Delta'<0$) in red.}
        \label{fig:AparTearTwist}
	\end{center}
\end{figure}
\begin{figure*}[htp]
	\begin{center}
	\includegraphics[width=\linewidth,keepaspectratio]{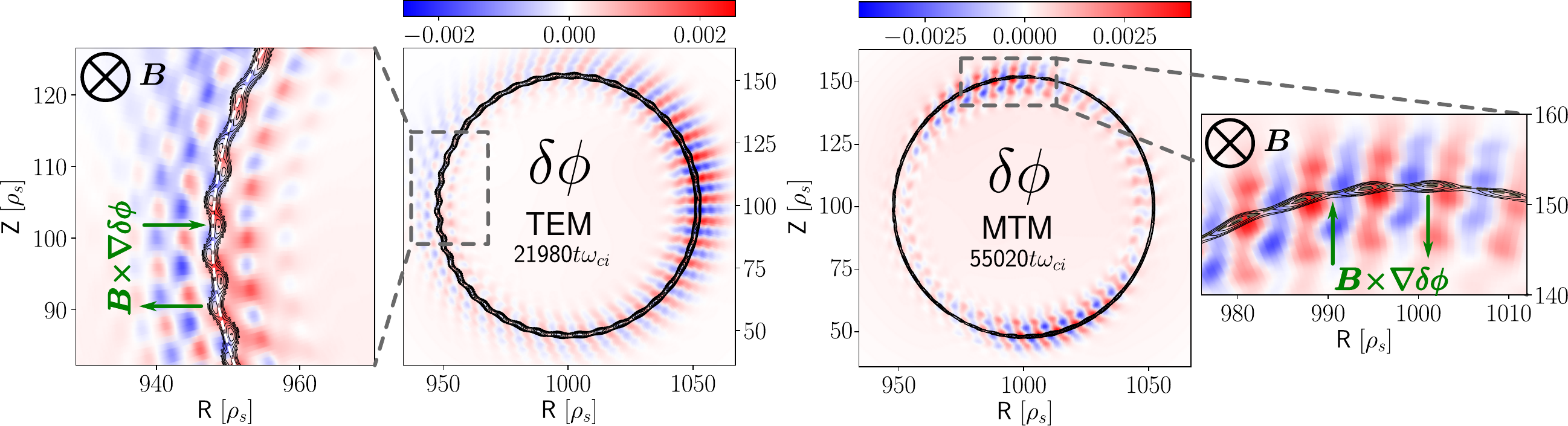}
        \caption{\justifying Poloidal cross-section of the perturbed electrostatic potential $\delta\phi$ during the exponential
                 growing phase of a TEM ($R/L_{T_s}=0$, $R/L_n=5$, $\beta=0.2\%$ and $m_i/m_e=200$), and an MTM
                 ($R/L_{T_i}=R/L_n=0$, $-R/L_{T_e}=9$, $\beta=0.05\%$ and $m_i/m_e=200$) simulations. The colorbars 
                  indicate the values of $\delta\phi$. The perturbed
                 $\cross{E}{B}=\cross{B}{\nabla\delta\phi}$ flows force the field line to reconnect and small-scale
                 islands form.}
        \label{fig:Tear_Twist}
	\end{center}
\end{figure*}
\newline\indent We employ the electromagnetic version of the Lagrangian gyrokinetic Particle-In-Cell (PIC) code ORB5 \cite{LantiCPC20} to investigate
whether turbulence can induce the growth of a large-scale magnetic island from a linearly stable tearing mode ($\Delta' < 0$). In ORB5, the equations are
solved using a control variate mitigation technique and a Pullback algorithm for the mixed-variables.\cite{MishchenkoCPC19}
While gyrokinetics is the most realistic framework for capturing micro-turbulence and MHD-scale interactions, we perform simulations
with an artificially large electron mass ratio ($m_i/m_e = 1/0.005$) to reduce computational costs and thus allow more extensive
parameter scans to investigate the underlying physics. The simulations are conducted in a circular, large-aspect-ratio geometry with $R_0/a = 10$,
where $a = 1$ m is the minor radius and $R_0 = 10$ m the major radius at the plasma center.
The on-axis magnetic field is $B_0 = 1$ T, and collisions are neglected. The ion sound Larmor radius is $\rho_s=\sqrt{T_e/m_i}=2a/L_x$ with
$L_x=200$, $\rho_i=\sqrt{2}\rho_s$ and the ion to electron temperature ratio is $T_i/T_e=1$. For these parameters, we are able to drive a reconnecting
instability, the micro-tearing mode (MTM),\cite{Drake77,IshizawaJPP15,HamedPoP19,XiePoP23} by setting a finite electron temperature gradient while imposing
flat density and ion-temperature profiles.
Conversely, a finite density or ion-temperature gradient leads to the destabilization of a
trapped-electron-mode (TEM)\cite{RyterPRL05,DannertPoP2005,Nakata2017} or an ion-temperature-gradient mode (ITG) respectively,\cite{Maeyama_PoP2014}
both characterized by twisting (or kinking) magnetic perturbations.\cite{IshizawaJPP15}
\newline\indent The grid resolution is ($n_s$,$n_\phi=128$,$n_\chi=384$), where $n_s$
is the radial direction, $n_\phi$ and $n_\chi$ are the toroidal and poloidal directions on which a Fourier decomposition is applied. The
number of toroidal and poloidal modes are $n\in[0,30]$ and $m\in[nq-5,nq+5]$. We select $n_s\in[500,1224]$ such that the
electron skin-depth $d_e/\rho_s=\sqrt{m_e/(m_i\beta_e)}$ is resolved by at least $7$ grid points. The number of markers for both electrons
and ions is $4\cdot10^8$. For a shifted electron distribution function (Maxwellian), the selected safety factor profile
$q=q_a(r/a)^2/(1-(1-r^2/a^2)^2)\label{eq:Safety}$, with $q_0=1.75$ and $q_a=q(r=a)=2q_0$ the on-axis and edge values,
produces a background current density $\boldsymbol{J}_0$. While this current can drive an unstable tearing mode ($\Delta'>0$) at the $q=2$
resonant surface (mid-radius),\cite{WidmerPoP24} we center the Maxwellian around zero, enforcing $\boldsymbol{J}_0\equiv 0$ at all times,
to obtain a linearly stable tearing mode ($\Delta'<0$).
\newline\indent Anticipating the results, \fref{fig:AparTearTwist} shows
the shape of the eigenfunction of the vector potential component parallel to the equilibrium magnetic field $(A_{||})$
in toroidal geometry for current- and turbulence-driven islands, the latter follows the vacuum structure.\cite{ZohmBook2022}
In presence of turbulence, magnetic reconnection occurs first on the small scales characterizing the most unstable mode that
drives the turbulence. We demonstrate below that for a given micro-instability with dominant toroidal and poloidal mode numbers $n$ and $m$,
approximately $m$ small-scale magnetic islands form at the rational surface, regardless of the driving mode parity.
\begin{figure}[htbp]
	\begin{center}
        \begin{subfigure}[t]{\linewidth}
	\begin{center}
	\includegraphics[width=0.6\linewidth,keepaspectratio]{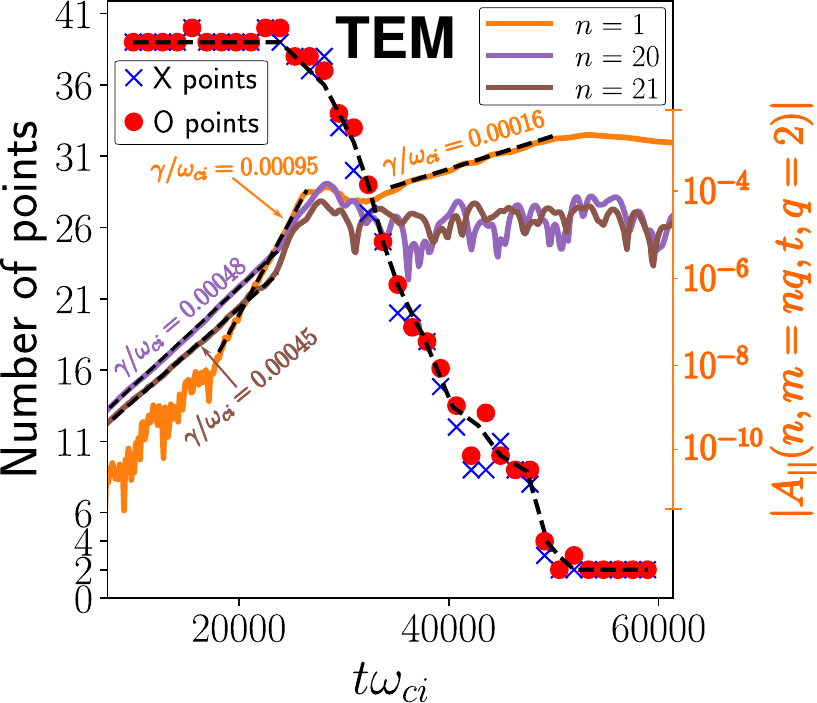}
        \caption{TEM case, $-R/L_n=5$, $-R/L_{T_s}=0$.}
        \label{fig:MTM_XOTime}
	\end{center}
        \end{subfigure}
        \begin{subfigure}[t]{\linewidth}
	\begin{center}
	\includegraphics[width=0.6\linewidth,keepaspectratio]{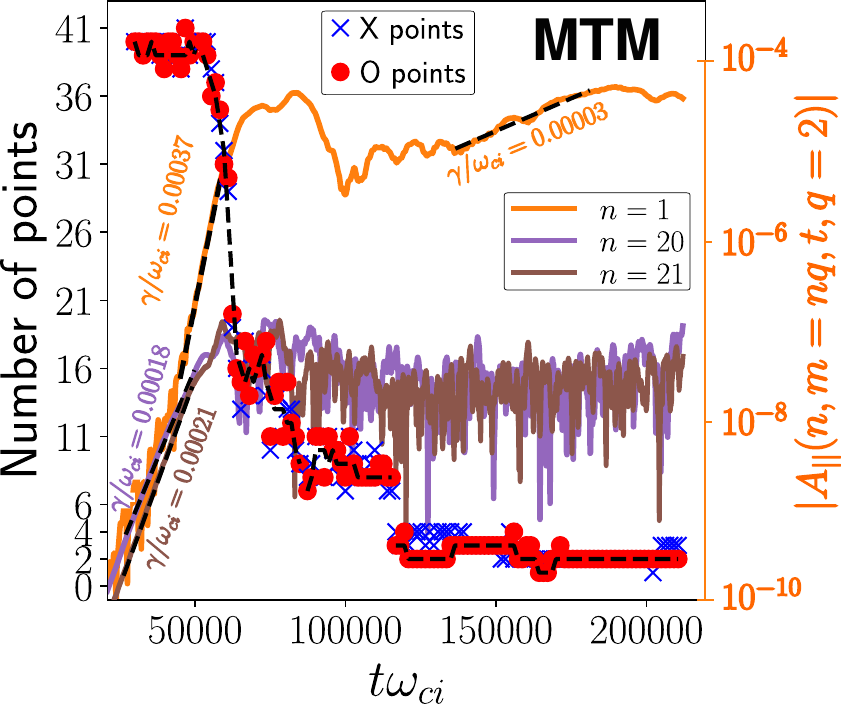}
        \caption{MTM case, $R/L_{T_i}=R/L_n=0$, $-R/L_{T_e}=9$. }
        \label{fig:ITG_XOTime}
	\end{center}
        \end{subfigure}
        \caption{\justifying Total number of X- and O-points along the resonant surface (blue crosses and red circles) alongside the temporal
                 evolution of the $|A_{||}|$ toroidal modes involved in the three-wave coupling for (a) TEM and (b) MTM, both with $\beta=0.05\%$
                 and $m_i/m_e=200$.}
        \label{fig:XOTime}
	\end{center}
\end{figure}
\newline\indent The poloidal cross sections of the
perturbed electrostatic potential $\delta\phi$ during the exponential-growth phase for tearing (MTM) and twisting parity (TEM) show the structure of the
small scales island (\fref{fig:Tear_Twist}). The perturbed $\cross{E}{B}$ flows are depicted in green. The island formation process exhibits distinct
characteristics in TEM and MTM. For the TEM, islands appear within the kinking perturbed magnetic surfaces,
similar to the scenario described in \citet{IgochinePoP14} and \citet{WillensdorferNatPhys24} where the perturbation is provided by a sawtooth crash
respectively by external coils. In contrast, the MTM exhibits symmetric reconnection around the rational surface. The formation of large-scale (low-$n$) magnetic islands embedded in reconnecting turbulent structures occurs in two phases,
as observed in \fref{fig:XOTime}.
\begin{figure}[htp]
     \centering
     \subfloat[TEM case I, $\beta$ scan.]{\includegraphics[width=0.3\textwidth,keepaspectratio]{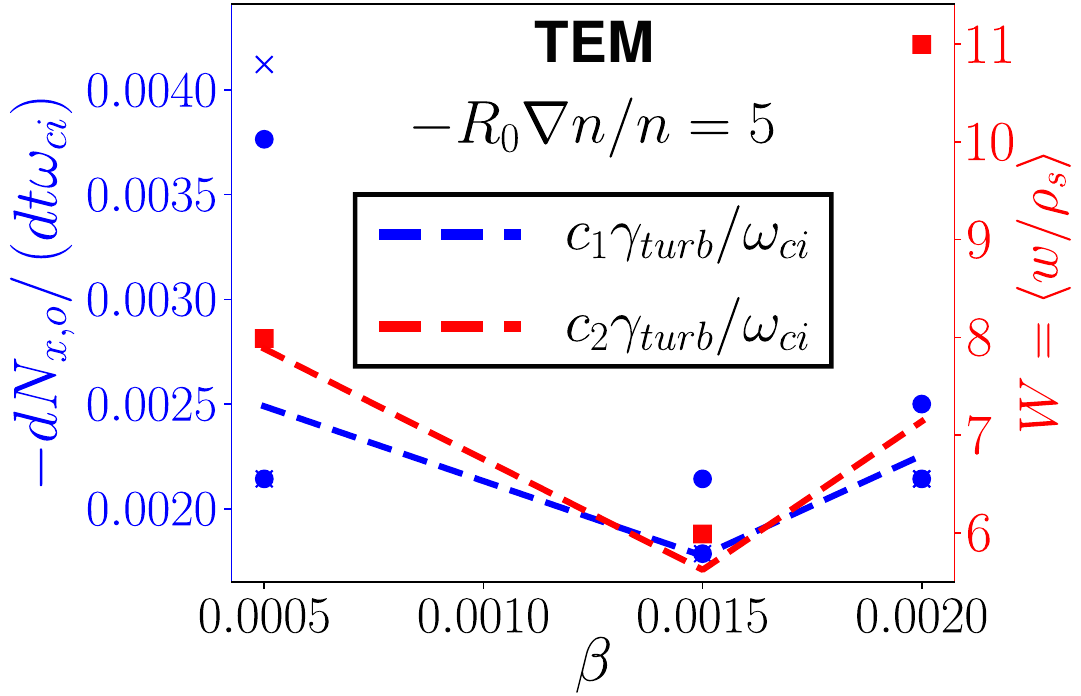}\label{subfig:TurbRecTEM5}}\\
     \subfloat[TEM case II, $\beta$ scan.]{\includegraphics[width=0.3\textwidth,keepaspectratio]{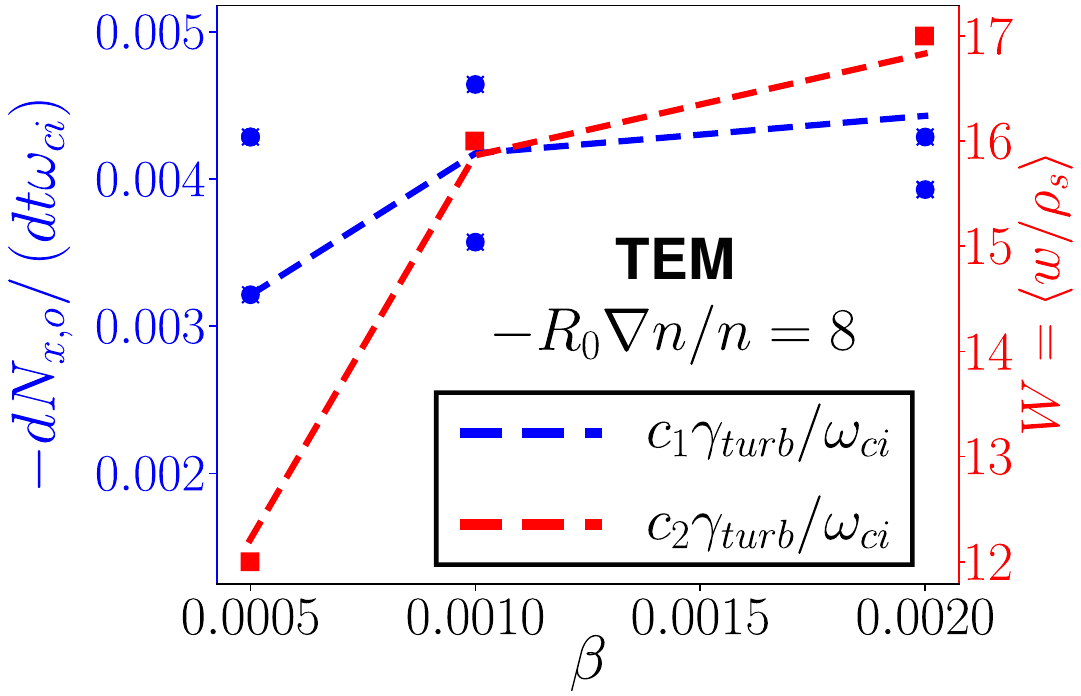}\label{subfig:TurbRecTEM8}}\\
     \subfloat[MTM case, $R/L_{T_e}$ scan.]{\includegraphics[width=0.3\textwidth,keepaspectratio]{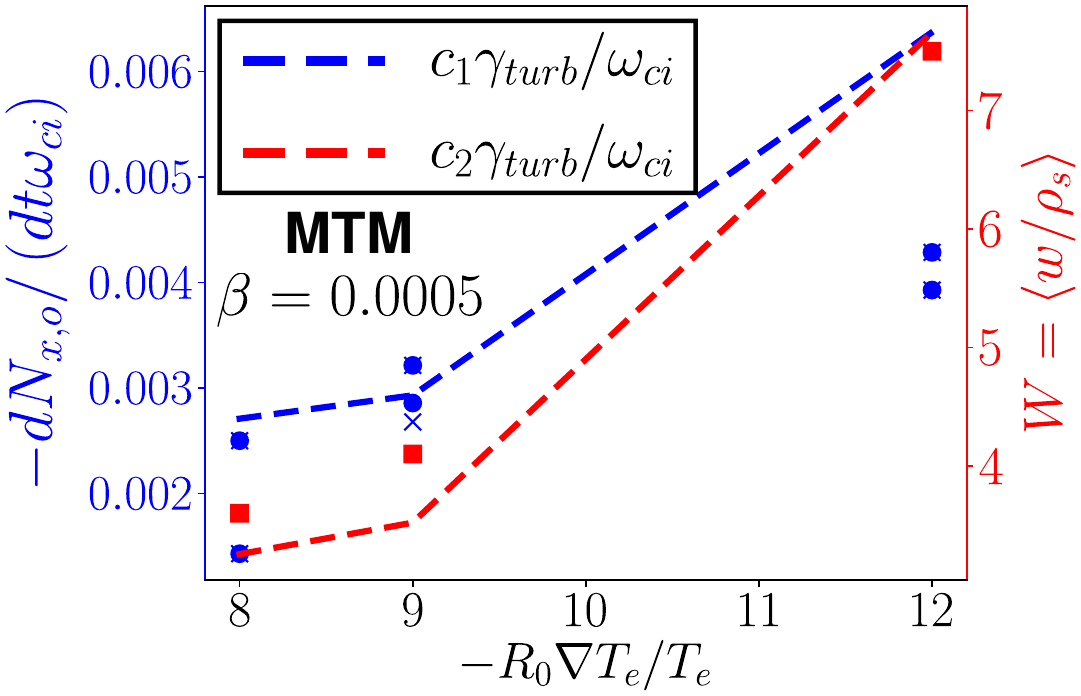}\label{subfig:TurbRecMTM}}
 \caption{\justifying Uncertainty extrema of the turbulent reconnection rate $-dN/(dt\omega_{ci})$, computed from the time-evolving
                 number of X/O points (blue crosses/circles, see \fref{fig:XOTime}), together with the measured average island width $W$ (red squares). Both quantities are
                 plotted versus the plasma beta for (a) and (b) (TEM cases) and (c) the electron temperature gradient
                 for the MTM case. The dashed-lines corresponds to \eref{eqn:gammaturb1} selecting a constant factor $C=c_1$ when they are compared with
                 $-dN/(dt\omega_{ci})$, and a constant factor $C=c_2$ when they are compared with $W$. The same values of $c_1$ and $c_2$ are used in all the
                 cases shown here.}
        \label{fig:TurbRec}
\end{figure}
\newline\indent The first phase occurs during the exponential growth of the relevant micro-instability and is characterized by a
beat-driven increase of the amplitude of long-wavelength components of $A_{||}$.
This mechanism was extensively investigated in the past, see \citet{MuragliaJPP25} and references therein.
Upon saturation of the micro-turbulence, a second phase emerges, characterized by the coalescence of small-scale islands into larger ones, accompanied by a
significant reduction in the number of X/O points (\fref{fig:XOTime}). While, the magnetic potential low-$n$ components experience a further order of magnitude increase, leading to the formation of a sizable magnetic
island, electrostatic potential components remain comparable in magnitude (not shown). This two-phase process, previously observed in fluid simulations,\cite{VillaPoP25} is reported here for the first time in the context of toroidal gyrokinetic simulations of tokamak plasmas.
\newline\indent The mechanism governing the coalescence is complex, yet independent of the mode parity. Based on the reduction in the number of X/O points $N \equiv m = nq_{rs}$, where $q_{rs}$ is the safety factor evaluated at the resonant surface, we quantify the coalescence by defining a normalized reconnection rate as $\gamma_{turb}(t)/\omega_{ci} = -dN/(dt\omega_{ci})$. We propose a dependence of
$\gamma_{turb}(t)/\omega_{ci}$ on the plasma beta, the micro-instability drive through the equilibrium profile logarithmic gradients, as well as the ratio
of electromagnetic to electrostatic fluctuations of the most unstable mode as
\begin{equation}
  \frac{\gamma_{turb}(t)}{\omega_{ci}} \approx C\left(\frac{|A_{||,n_{inst}}|}{|\phi_{n_{inst}}|}\right)^{1/2}\frac{d_e}{\rho_s}\left|R_0\frac{\nabla p}{p}\right|,\label{eqn:gammaturb1}
\end{equation}
where $C$ is a numerical constant, $n_{inst}=m_{inst}/q_{rs}$ is the dominant micro-instability toroidal mode number, $|A_{||,n_{inst}}|$ and $|\phi_{n_{inst}}|$ are its potentials
evaluated at saturation once the merging begins. \Eref{eqn:gammaturb1} highlights the importance of the initial thickness of the
reconnection layer, related to $d_e/\rho_s$, and the small-scale islands width proportional to $\sqrt{|A_{||,n_{inst}}|}$. \Fref{fig:TurbRec} shows that
our conjecture scales well with the computed $-dN/(dt\omega_{ci})$ from the simulation results, up to a constant $C=c_1$, for the $\beta$ (TEM) and $R/L_{T_e}$ (MTM) scans available. Furthermore, \eref{eqn:gammaturb1} scales with the averaged (long-wavelength) island width $W=\langle w/\rho_s\rangle$ using a constant $C=c_2$.
These findings suggest that the size of turbulence-driven magnetic islands is heavily influenced by the micro-instability drive and saturation mechanism.
\begin{figure}
	\begin{center}
	\includegraphics[width=0.7\linewidth,keepaspectratio]{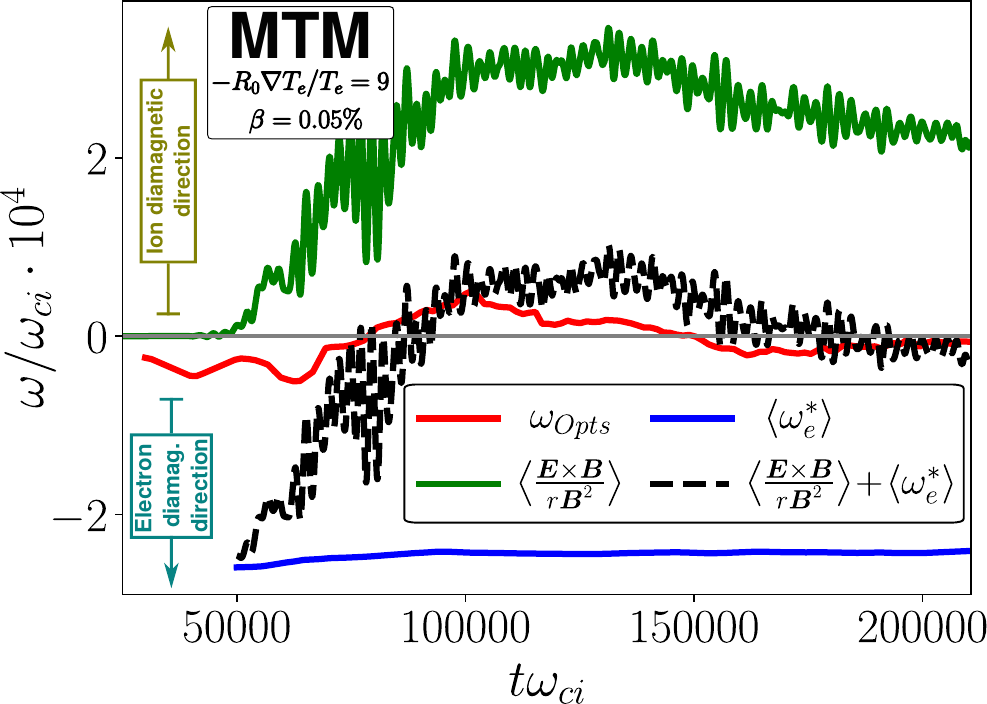}
        \end{center}
		\caption{\justifying Correlation between island rotation, zonal flow dynamics and electron diamagnetic
                         frequency $\omega^*_e=m\cross{B}{\nabla T_e}/(r_sB^2)$ (blue line) highlighted
                         using the rotation frequency of O-points $\omega_{Opts}$ (red line) and zonal flow
                         angular velocity $\cross{E}{B}/(B^2)$ (green line). The average $\langle\cdots\rangle$ is performed over
                         the large (long-wavelength) island width.                        }
        \label{fig:FreqMTM}
\end{figure}
\begin{figure*}
	\begin{center}
	\includegraphics[width=\linewidth,keepaspectratio]{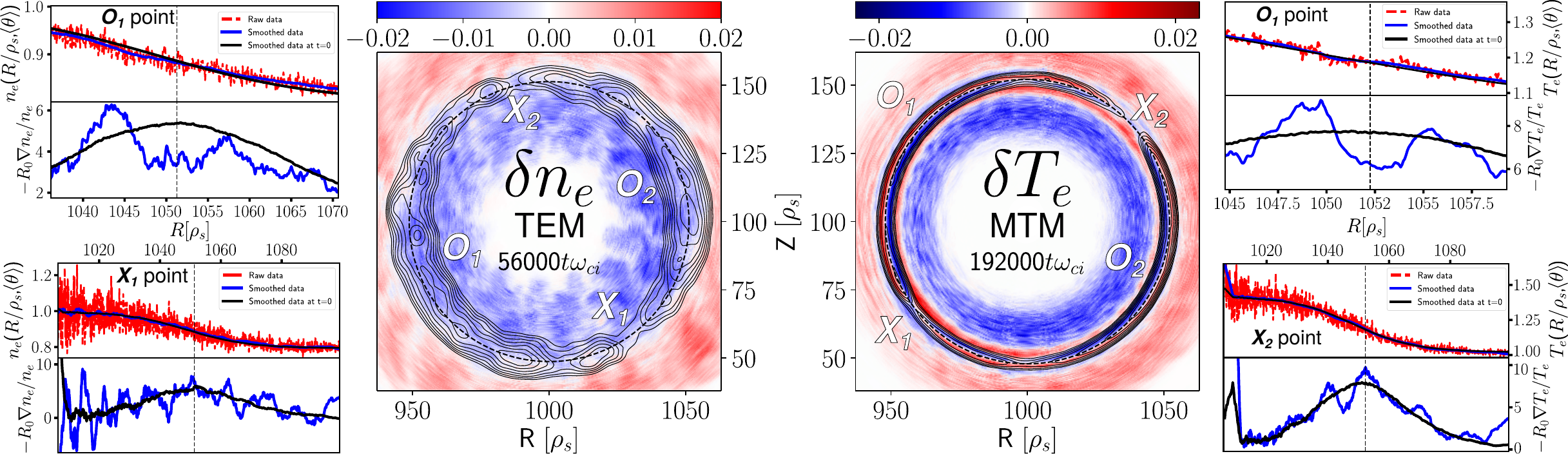}
	\caption{\justifying Poloidal cross-sections of TEM and MTM turbulence-induced islands during the stationary phase of the turbulence.
          The colorbars indicate the values of the electron density fluctuations $\delta n_e$ for the MTM and the electron temperature
          fluctuations $\delta T_e$ for the MTM.
          The plots on the sides show the profile and the logarithmic gradient of the electron density (left) and electron temperature (right) calculated
          by poloidally averaging across the O-point (top) and X-point (bottom). A decrease of the gradients around the O-point can be observed.}
        \label{fig:2dProfs}
	\end{center}
\end{figure*}
\newline\indent Given its importance in terms of neoclassical physics, see \citet{PoliPPR16} and references therein, we add here a brief remark concerning
the rotation of turbulence-induced magnetic islands, although a detailed analysis is outside the scope of this paper. As the island forms, its rotation undergoes a
significant change, transitioning from electron diamagnetic to ion diamagnetic (\fref{fig:FreqMTM}). This behaviour, closely related to the growth of the
zonal flow in the island region, is a common observation in toroidal simulations of coupled turbulence/tearing mode dynamics.\cite{IshizawaPPCF19}
The rotation reversal may be linked to the different nonlinear evolution of ion and electron temperature profiles across the island. In this scenario,
an electric field is predicted to develop, ensuring the rotation of the electron fluid with the island. \cite{SmolyakovPoP95,Wilson1996a} This prediction
has been supported by simulation results. \cite{SiccinioPoP11,WidmerPoP24}
\newline\indent The merging mechanism is a universal feature, independent of the mode parity of the underlying micro-instability, capable of nonlinearly forming
large-scale magnetic islands. The final island size depends, however, on the initial micro-instability drive, the electron skin-depth and the
amplitude of the electrostatic and magnetic potentials (\eref{eqn:gammaturb1}). For instance, the TEM-induced $2/1$-island exhibits several
sub-islands and an irregular separatrix, in contrast to the MTM-induced $2/1$-island, which shows fewer substructures and a smoother separatrix
(see \fref{fig:2dProfs}).
While these differences have likely an impact on the details of the transport in the island region, a common observation
emerging from a poloidally-resolved analysis of the profiles is that the gradients are significantly reduced across the O-point, while they remain close
to the equilibrium values when evaluated around the X-point, as clearly visible in the side panels of \fref{fig:2dProfs}. The radial profiles used to
compute the logarithmic gradients are first smoothed radially by a moving-average procedure. This is followed by a
poloidal average over the island extent for O-points and between two islands for X-points.
A similar analysis\footnote{for the simulation results reported in \fref{fig:MTMTeProfs}), poloidally-resolved profiles are unfortunately
not available; flux-surface-averaged profiles are shown instead.}
performed for ITG turbulence (\fref{fig:MTMTeProfs}) confirms that a significant flattening (as high as 50\% for the ion temperature
gradient with $\RLN{}=\RLT{e}=0$, $\RLT{i}=8$, $\beta=0.05\%$) can occur. This is due to the formation of a large-scale island and is independent of the details of its generation.
\newline\indent We have demonstrated, for the first time using the gyrokinetic formalism, that large-scale magnetic islands can emerge in plasmas that are
linearly stable to tearing modes ($\Delta' < 0$). This phenomenon occurs through the induction of turbulence by micro-instabilities. Our findings show that
small-scale magnetic islands form at the rational surface from perturbed $\cross{E}{B}$ flows, with their number characterized by the dominant
instability poloidal mode number. In the presence of twisting parity instabilities, these islands are embedded within kinked perturbed flux surfaces.
Regardless of the mode parity, we identified two phases in the formation of the large scale islands. The (linearly stable) $n=1$ mode first grows from
three-wave-coupling with large-$n$ modes. At their saturation, the small-scale islands then coalesce rapidly, building large-scale $2/1$-magnetic
islands. We have identified the coalescence of small-scale magnetic islands as a universal phenomenon, which we quantified by defining a turbulent
reconnection rate. We obtained that the turbulence-driven $2/1$-magnetic island final width, calculated from simulation results, is proportional to our
turbulent reconnection rate, suggesting a strong influence of the micro-instability drive and saturation mechanism.
The overall process produces a $2/1$-magnetic island, with a rotation frequency correlated with the zonal flow dynamics, that is large enough to
significantly flatten the pressure profile. Such islands can therefore be regarded as a seed for the NTM onset.
\section{Acknowledgments}
This work was carried out using supercomputer resources of JFRS-1 provided under the EU-JA Broader Approach collaboration in the Computational Simulation Centre of International Fusion Energy Research Centre (IFERC-CSC) and on the MARCONI FUSION HPC system at CINECA. Furthermore, part of this work has been carried out within the framework of the EUROfusion Consortium, funded by the European Union via the Euratom Research and Training Programme (Grant Agreement No. 101052200—EUROfusion). Views and opinions expressed are however those of the authors only and do not necessarily reflect those of the European Union or the European Commission. Neither the European Union nor the European Commission can be held responsible for them. Part of this work was done when FW was employed by National Institutes of Natural Sciences (NINS), Tokyo, Japan. EP would like to thank the
Graduate School of Energy Science of the Kyoto University for its hospitality during the early phase of this study.
EP is partially supported by the Eurofusion Enabling Research Project (CfP-FSD-AWP24-ENR T-RecS). AB, AM, and THS are partially supported by Eurofusion Theory and Simulation Verification and Validation Task 10. FW and EP wish to thank V. Igochine and C. Angioni for valuable discussions. Finally, the authors are grateful for the numerous discussions and suggestions within the framework of both EUROfusion projects.
\bibliography{library}

\begin{thebibliography}{48}%
\makeatletter
\providecommand \@ifxundefined [1]{%
 \@ifx{#1\undefined}
}%
\providecommand \@ifnum [1]{%
 \ifnum #1\expandafter \@firstoftwo
 \else \expandafter \@secondoftwo
 \fi
}%
\providecommand \@ifx [1]{%
 \ifx #1\expandafter \@firstoftwo
 \else \expandafter \@secondoftwo
 \fi
}%
\providecommand \natexlab [1]{#1}%
\providecommand \enquote  [1]{``#1''}%
\providecommand \bibnamefont  [1]{#1}%
\providecommand \bibfnamefont [1]{#1}%
\providecommand \citenamefont [1]{#1}%
\providecommand \href@noop [0]{\@secondoftwo}%
\providecommand \href [0]{\begingroup \@sanitize@url \@href}%
\providecommand \@href[1]{\@@startlink{#1}\@@href}%
\providecommand \@@href[1]{\endgroup#1\@@endlink}%
\providecommand \@sanitize@url [0]{\catcode `\\12\catcode `\$12\catcode
  `\&12\catcode `\#12\catcode `\^12\catcode `\_12\catcode `\%12\relax}%
\providecommand \@@startlink[1]{}%
\providecommand \@@endlink[0]{}%
\providecommand \url  [0]{\begingroup\@sanitize@url \@url }%
\providecommand \@url [1]{\endgroup\@href {#1}{\urlprefix }}%
\providecommand \urlprefix  [0]{URL }%
\providecommand \Eprint [0]{\href }%
\providecommand \doibase [0]{https://doi.org/}%
\providecommand \selectlanguage [0]{\@gobble}%
\providecommand \bibinfo  [0]{\@secondoftwo}%
\providecommand \bibfield  [0]{\@secondoftwo}%
\providecommand \translation [1]{[#1]}%
\providecommand \BibitemOpen [0]{}%
\providecommand \bibitemStop [0]{}%
\providecommand \bibitemNoStop [0]{.\EOS\space}%
\providecommand \EOS [0]{\spacefactor3000\relax}%
\providecommand \BibitemShut  [1]{\csname bibitem#1\endcsname}%
\let\auto@bib@innerbib\@empty
\bibitem [{\citenamefont {{Ida}}\ \emph {et~al.}(2018)\citenamefont {{Ida}},
  \citenamefont {{Kobayashi}}, \citenamefont {{Ono}}, \citenamefont {{Evans}},
  \citenamefont {{McKee}},\ and\ \citenamefont {{Austin}}}]{IdaPRL18}%
  \BibitemOpen
  \bibfield  {author} {\bibinfo {author} {\bibfnamefont {K.}~\bibnamefont
  {{Ida}}}, \bibinfo {author} {\bibfnamefont {T.}~\bibnamefont {{Kobayashi}}},
  \bibinfo {author} {\bibfnamefont {M.}~\bibnamefont {{Ono}}}, \bibinfo
  {author} {\bibfnamefont {T.~E.}\ \bibnamefont {{Evans}}}, \bibinfo {author}
  {\bibfnamefont {G.~R.}\ \bibnamefont {{McKee}}},\ and\ \bibinfo {author}
  {\bibfnamefont {M.~E.}\ \bibnamefont {{Austin}}},\ }\bibfield  {title}
  {\enquote {\bibinfo {title} {{Hysteresis Relation between Turbulence and
  Temperature Modulation during the Heat Pulse Propagation into a Magnetic
  Island in DIII-D}},}\ }\href {https://doi.org/10.1103/PhysRevLett.120.245001}
  {\bibfield  {journal} {\bibinfo  {journal} {prl}\ }\textbf {\bibinfo {volume}
  {120}},\ \bibinfo {eid} {245001} (\bibinfo {year} {2018})}\BibitemShut
  {NoStop}%
\bibitem [{\citenamefont {{Bard{\'o}czi}}\ and\ \citenamefont
  {{Evans}}(2021)}]{BardocziPRL21}%
  \BibitemOpen
  \bibfield  {author} {\bibinfo {author} {\bibfnamefont {L.}~\bibnamefont
  {{Bard{\'o}czi}}}\ and\ \bibinfo {author} {\bibfnamefont {T.~E.}\
  \bibnamefont {{Evans}}},\ }\bibfield  {title} {\enquote {\bibinfo {title}
  {{Experimental Observation of Magnetic Island Heteroclinic Bifurcation in
  Tokamaks}},}\ }\href {https://doi.org/10.1103/PhysRevLett.126.085003}
  {\bibfield  {journal} {\bibinfo  {journal} {prl}\ }\textbf {\bibinfo {volume}
  {126}},\ \bibinfo {eid} {085003} (\bibinfo {year} {2021})}\BibitemShut
  {NoStop}%
\bibitem [{\citenamefont {{Bardoczi}}\ \emph {et~al.}(2024)\citenamefont
  {{Bardoczi}}, \citenamefont {{Richner}}, \citenamefont {{Logan}},
  \citenamefont {{Strait}}, \citenamefont {{Holcomb}}, \citenamefont {{Zhu}},\
  and\ \citenamefont {{Rea}}}]{BardocziNF24}%
  \BibitemOpen
  \bibfield  {author} {\bibinfo {author} {\bibfnamefont {L.}~\bibnamefont
  {{Bardoczi}}}, \bibinfo {author} {\bibfnamefont {N.~J.}\ \bibnamefont
  {{Richner}}}, \bibinfo {author} {\bibfnamefont {N.~C.}\ \bibnamefont
  {{Logan}}}, \bibinfo {author} {\bibfnamefont {E.~J.}\ \bibnamefont
  {{Strait}}}, \bibinfo {author} {\bibfnamefont {C.~T.}\ \bibnamefont
  {{Holcomb}}}, \bibinfo {author} {\bibfnamefont {J.}~\bibnamefont {{Zhu}}},\
  and\ \bibinfo {author} {\bibfnamefont {C.}~\bibnamefont {{Rea}}},\ }\bibfield
   {title} {\enquote {\bibinfo {title} {{The root cause of disruptive NTMs and
  paths to stable operation in DIII-D ITER baseline scenario plasmas}},}\
  }\href {https://doi.org/10.1088/1741-4326/ad7787} {\bibfield  {journal}
  {\bibinfo  {journal} {Nuclear Fusion}\ }\textbf {\bibinfo {volume} {64}},\
  \bibinfo {eid} {126005} (\bibinfo {year} {2024})}\BibitemShut {NoStop}%
\bibitem [{\citenamefont {Kotschenreuther}, \citenamefont {Hazeltine},\ and\
  \citenamefont {Morrison}(1985)}]{kotschenreuther_PF1985}%
  \BibitemOpen
  \bibfield  {author} {\bibinfo {author} {\bibfnamefont {M.}~\bibnamefont
  {Kotschenreuther}}, \bibinfo {author} {\bibfnamefont {R.~D.}\ \bibnamefont
  {Hazeltine}},\ and\ \bibinfo {author} {\bibfnamefont {P.~J.}\ \bibnamefont
  {Morrison}},\ }\bibfield  {title} {\enquote {\bibinfo {title} {{Nonlinear
  dynamics of magnetic islands with curvature and pressure}},}\ }\href
  {https://doi.org/10.1063/1.865200} {\bibfield  {journal} {\bibinfo  {journal}
  {Physics of Fluids}\ }\textbf {\bibinfo {volume} {28}},\ \bibinfo {pages}
  {294--302} (\bibinfo {year} {1985})}\BibitemShut {NoStop}%
\bibitem [{\citenamefont {Carreras}, \citenamefont {Hicks},\ and\ \citenamefont
  {Lee}(1981)}]{Carreras1981a}%
  \BibitemOpen
  \bibfield  {author} {\bibinfo {author} {\bibfnamefont {B.}~\bibnamefont
  {Carreras}}, \bibinfo {author} {\bibfnamefont {H.~R.}\ \bibnamefont
  {Hicks}},\ and\ \bibinfo {author} {\bibfnamefont {D.~K.}\ \bibnamefont
  {Lee}},\ }\bibfield  {title} {\enquote {\bibinfo {title} {{Effects of
  toroidal coupling on the stability of tearing modes}},}\ }\href
  {https://doi.org/10.1063/1.863247} {\bibfield  {journal} {\bibinfo  {journal}
  {Physics of Fluids}\ }\textbf {\bibinfo {volume} {24}},\ \bibinfo {pages}
  {66--77} (\bibinfo {year} {1981})}\BibitemShut {NoStop}%
\bibitem [{\citenamefont {Sauter}\ \emph {et~al.}(1997)\citenamefont {Sauter},
  \citenamefont {{La Haye}}, \citenamefont {Chang}, \citenamefont {Gates},
  \citenamefont {Kamada}, \citenamefont {Zohm}, \citenamefont {Bondeson},
  \citenamefont {Boucher}, \citenamefont {Callen}, \citenamefont {Chu},
  \citenamefont {Gianakon}, \citenamefont {Gruber}, \citenamefont {Harvey},
  \citenamefont {Hegna}, \citenamefont {Lao}, \citenamefont {Monticello},
  \citenamefont {Perkins}, \citenamefont {Pletzer}, \citenamefont {Reiman},
  \citenamefont {Rosenbluth}, \citenamefont {Strait}, \citenamefont {Taylor},
  \citenamefont {Turnbull}, \citenamefont {Waelbroeck}, \citenamefont {Wesley},
  \citenamefont {Wilson},\ and\ \citenamefont {Yoshino}}]{SauterPoP97a}%
  \BibitemOpen
  \bibfield  {author} {\bibinfo {author} {\bibfnamefont {O.}~\bibnamefont
  {Sauter}}, \bibinfo {author} {\bibfnamefont {R.~J.}\ \bibnamefont {{La
  Haye}}}, \bibinfo {author} {\bibfnamefont {Z.}~\bibnamefont {Chang}},
  \bibinfo {author} {\bibfnamefont {D.~A.}\ \bibnamefont {Gates}}, \bibinfo
  {author} {\bibfnamefont {Y.}~\bibnamefont {Kamada}}, \bibinfo {author}
  {\bibfnamefont {H.}~\bibnamefont {Zohm}}, \bibinfo {author} {\bibfnamefont
  {A.}~\bibnamefont {Bondeson}}, \bibinfo {author} {\bibfnamefont
  {D.}~\bibnamefont {Boucher}}, \bibinfo {author} {\bibfnamefont {J.~D.}\
  \bibnamefont {Callen}}, \bibinfo {author} {\bibfnamefont {M.~S.}\
  \bibnamefont {Chu}}, \bibinfo {author} {\bibfnamefont {T.~A.}\ \bibnamefont
  {Gianakon}}, \bibinfo {author} {\bibfnamefont {O.}~\bibnamefont {Gruber}},
  \bibinfo {author} {\bibfnamefont {R.~W.}\ \bibnamefont {Harvey}}, \bibinfo
  {author} {\bibfnamefont {C.~C.}\ \bibnamefont {Hegna}}, \bibinfo {author}
  {\bibfnamefont {L.~L.}\ \bibnamefont {Lao}}, \bibinfo {author} {\bibfnamefont
  {D.~A.}\ \bibnamefont {Monticello}}, \bibinfo {author} {\bibfnamefont
  {F.}~\bibnamefont {Perkins}}, \bibinfo {author} {\bibfnamefont
  {A.}~\bibnamefont {Pletzer}}, \bibinfo {author} {\bibfnamefont {A.~H.}\
  \bibnamefont {Reiman}}, \bibinfo {author} {\bibfnamefont {M.}~\bibnamefont
  {Rosenbluth}}, \bibinfo {author} {\bibfnamefont {E.~J.}\ \bibnamefont
  {Strait}}, \bibinfo {author} {\bibfnamefont {T.~S.}\ \bibnamefont {Taylor}},
  \bibinfo {author} {\bibfnamefont {A.~D.}\ \bibnamefont {Turnbull}}, \bibinfo
  {author} {\bibfnamefont {F.}~\bibnamefont {Waelbroeck}}, \bibinfo {author}
  {\bibfnamefont {J.~C.}\ \bibnamefont {Wesley}}, \bibinfo {author}
  {\bibfnamefont {H.~R.}\ \bibnamefont {Wilson}},\ and\ \bibinfo {author}
  {\bibfnamefont {R.}~\bibnamefont {Yoshino}},\ }\bibfield  {title} {\enquote
  {\bibinfo {title} {{Beta limits in long-pulse tokamak discharges}},}\ }\href
  {https://doi.org/10.1063/1.872270} {\bibfield  {journal} {\bibinfo  {journal}
  {Physics of Plasmas}\ }\textbf {\bibinfo {volume} {4}},\ \bibinfo {pages}
  {1654--1664} (\bibinfo {year} {1997})}\BibitemShut {NoStop}%
\bibitem [{\citenamefont {{Sweeney}}\ \emph {et~al.}(2017)\citenamefont
  {{Sweeney}}, \citenamefont {{Choi}}, \citenamefont {{La Haye}}, \citenamefont
  {{Mao}}, \citenamefont {{Olofsson}}, \citenamefont {{Volpe}},\ and\
  \citenamefont {{The DIII-D Team}}}]{SweeneyNF17}%
  \BibitemOpen
  \bibfield  {author} {\bibinfo {author} {\bibfnamefont {R.}~\bibnamefont
  {{Sweeney}}}, \bibinfo {author} {\bibfnamefont {W.}~\bibnamefont {{Choi}}},
  \bibinfo {author} {\bibfnamefont {R.~J.}\ \bibnamefont {{La Haye}}}, \bibinfo
  {author} {\bibfnamefont {S.}~\bibnamefont {{Mao}}}, \bibinfo {author}
  {\bibfnamefont {K.~E.~J.}\ \bibnamefont {{Olofsson}}}, \bibinfo {author}
  {\bibfnamefont {F.~A.}\ \bibnamefont {{Volpe}}},\ and\ \bibinfo {author}
  {\bibnamefont {{The DIII-D Team}}},\ }\bibfield  {title} {\enquote {\bibinfo
  {title} {{Statistical analysis of m/n{\,}{\,}={\,}{\,}2/1 locked and
  quasi-stationary modes with rotating precursors at DIII-D}},}\ }\href
  {https://doi.org/10.1088/0029-5515/57/1/016019} {\bibfield  {journal}
  {\bibinfo  {journal} {Nuclear Fusion}\ }\textbf {\bibinfo {volume} {57}},\
  \bibinfo {eid} {016019} (\bibinfo {year} {2017})},\ \Eprint
  {https://arxiv.org/abs/1606.04183} {arXiv:1606.04183 [physics.plasm-ph]}
  \BibitemShut {NoStop}%
\bibitem [{\citenamefont {{Zohm}}\ \emph {et~al.}(1999)\citenamefont {{Zohm}},
  \citenamefont {{Gantenbein}}, \citenamefont {{Giruzzi}}, \citenamefont
  {{G{\"u}nter}}, \citenamefont {{Leuterer}}, \citenamefont {{Maraschek}},
  \citenamefont {{Meskat}}, \citenamefont {{Peeters}}, \citenamefont
  {{Suttrop}}, \citenamefont {{Wagner}}, \citenamefont {{Zabi{\'e}go}},
  \citenamefont {{ASDEX Upgrade Team}},\ and\ \citenamefont {{ECRH
  Group}}}]{ZohmNF99}%
  \BibitemOpen
  \bibfield  {author} {\bibinfo {author} {\bibfnamefont {H.}~\bibnamefont
  {{Zohm}}}, \bibinfo {author} {\bibfnamefont {G.}~\bibnamefont
  {{Gantenbein}}}, \bibinfo {author} {\bibfnamefont {G.}~\bibnamefont
  {{Giruzzi}}}, \bibinfo {author} {\bibfnamefont {S.}~\bibnamefont
  {{G{\"u}nter}}}, \bibinfo {author} {\bibfnamefont {F.}~\bibnamefont
  {{Leuterer}}}, \bibinfo {author} {\bibfnamefont {M.}~\bibnamefont
  {{Maraschek}}}, \bibinfo {author} {\bibfnamefont {J.}~\bibnamefont
  {{Meskat}}}, \bibinfo {author} {\bibfnamefont {A.~G.}\ \bibnamefont
  {{Peeters}}}, \bibinfo {author} {\bibfnamefont {W.}~\bibnamefont
  {{Suttrop}}}, \bibinfo {author} {\bibfnamefont {D.}~\bibnamefont {{Wagner}}},
  \bibinfo {author} {\bibfnamefont {M.}~\bibnamefont {{Zabi{\'e}go}}}, \bibinfo
  {author} {\bibnamefont {{ASDEX Upgrade Team}}},\ and\ \bibinfo {author}
  {\bibnamefont {{ECRH Group}}},\ }\bibfield  {title} {\enquote {\bibinfo
  {title} {{LETTER: Experiments on neoclassical tearing mode stabilization by
  ECCD in ASDEX Upgrade}},}\ }\href
  {https://doi.org/10.1088/0029-5515/39/5/101} {\bibfield  {journal} {\bibinfo
  {journal} {Nuclear Fusion}\ }\textbf {\bibinfo {volume} {39}},\ \bibinfo
  {pages} {577--580} (\bibinfo {year} {1999})}\BibitemShut {NoStop}%
\bibitem [{\citenamefont {{Prater}}\ \emph {et~al.}(2007)\citenamefont
  {{Prater}}, \citenamefont {{La Haye}}, \citenamefont {{Luce}}, \citenamefont
  {{Petty}}, \citenamefont {{Strait}}, \citenamefont {{Ferron}}, \citenamefont
  {{Humphreys}}, \citenamefont {{Isayama}}, \citenamefont {{Lohr}},
  \citenamefont {{Nagasaki}}, \citenamefont {{Politzer}}, \citenamefont
  {{Wade}},\ and\ \citenamefont {{Welander}}}]{PraterNF07}%
  \BibitemOpen
  \bibfield  {author} {\bibinfo {author} {\bibfnamefont {R.}~\bibnamefont
  {{Prater}}}, \bibinfo {author} {\bibfnamefont {R.~J.}\ \bibnamefont {{La
  Haye}}}, \bibinfo {author} {\bibfnamefont {T.~C.}\ \bibnamefont {{Luce}}},
  \bibinfo {author} {\bibfnamefont {C.~C.}\ \bibnamefont {{Petty}}}, \bibinfo
  {author} {\bibfnamefont {E.~J.}\ \bibnamefont {{Strait}}}, \bibinfo {author}
  {\bibfnamefont {J.~R.}\ \bibnamefont {{Ferron}}}, \bibinfo {author}
  {\bibfnamefont {D.~A.}\ \bibnamefont {{Humphreys}}}, \bibinfo {author}
  {\bibfnamefont {A.}~\bibnamefont {{Isayama}}}, \bibinfo {author}
  {\bibfnamefont {J.}~\bibnamefont {{Lohr}}}, \bibinfo {author} {\bibfnamefont
  {K.}~\bibnamefont {{Nagasaki}}}, \bibinfo {author} {\bibfnamefont {P.~A.}\
  \bibnamefont {{Politzer}}}, \bibinfo {author} {\bibfnamefont {M.~R.}\
  \bibnamefont {{Wade}}},\ and\ \bibinfo {author} {\bibfnamefont {A.~S.}\
  \bibnamefont {{Welander}}},\ }\bibfield  {title} {\enquote {\bibinfo {title}
  {{Stabilization and prevention of the 2/1 neoclassical tearing mode for
  improved performance in DIII-D}},}\ }\href
  {https://doi.org/10.1088/0029-5515/47/5/001} {\bibfield  {journal} {\bibinfo
  {journal} {Nuclear Fusion}\ }\textbf {\bibinfo {volume} {47}},\ \bibinfo
  {pages} {371--377} (\bibinfo {year} {2007})}\BibitemShut {NoStop}%
\bibitem [{\citenamefont {Isayama}\ \emph {et~al.}(2009)\citenamefont
  {Isayama}, \citenamefont {Matsunaga}, \citenamefont {Kobayashi},
  \citenamefont {Moriyama}, \citenamefont {Oyama}, \citenamefont {Sakamoto},
  \citenamefont {Suzuki}, \citenamefont {Urano}, \citenamefont {Hayashi},
  \citenamefont {Kamada}, \citenamefont {Ozeki}, \citenamefont {Hirano},
  \citenamefont {Urso}, \citenamefont {Zohm}, \citenamefont {Maraschek},
  \citenamefont {Hobirk}, \citenamefont {Nagasaki},\ and\ \citenamefont {the
  JT-60~team}}]{Isayama_NF2009}%
  \BibitemOpen
  \bibfield  {author} {\bibinfo {author} {\bibfnamefont {A.}~\bibnamefont
  {Isayama}}, \bibinfo {author} {\bibfnamefont {G.}~\bibnamefont {Matsunaga}},
  \bibinfo {author} {\bibfnamefont {T.}~\bibnamefont {Kobayashi}}, \bibinfo
  {author} {\bibfnamefont {S.}~\bibnamefont {Moriyama}}, \bibinfo {author}
  {\bibfnamefont {N.}~\bibnamefont {Oyama}}, \bibinfo {author} {\bibfnamefont
  {Y.}~\bibnamefont {Sakamoto}}, \bibinfo {author} {\bibfnamefont
  {T.}~\bibnamefont {Suzuki}}, \bibinfo {author} {\bibfnamefont
  {H.}~\bibnamefont {Urano}}, \bibinfo {author} {\bibfnamefont
  {N.}~\bibnamefont {Hayashi}}, \bibinfo {author} {\bibfnamefont
  {Y.}~\bibnamefont {Kamada}}, \bibinfo {author} {\bibfnamefont
  {T.}~\bibnamefont {Ozeki}}, \bibinfo {author} {\bibfnamefont
  {Y.}~\bibnamefont {Hirano}}, \bibinfo {author} {\bibfnamefont
  {L.}~\bibnamefont {Urso}}, \bibinfo {author} {\bibfnamefont {H.}~\bibnamefont
  {Zohm}}, \bibinfo {author} {\bibfnamefont {M.}~\bibnamefont {Maraschek}},
  \bibinfo {author} {\bibfnamefont {J.}~\bibnamefont {Hobirk}}, \bibinfo
  {author} {\bibfnamefont {K.}~\bibnamefont {Nagasaki}},\ and\ \bibinfo
  {author} {\bibnamefont {the JT-60~team}},\ }\bibfield  {title} {\enquote
  {\bibinfo {title} {{Neoclassical tearing mode control using electron
  cyclotron current drive and magnetic island evolution in {\{}JT-60U{\}}}},}\
  }\href {https://doi.org/10.1088/0029-5515/49/5/055006} {\bibfield  {journal}
  {\bibinfo  {journal} {Nuclear Fusion}\ }\textbf {\bibinfo {volume} {49}},\
  \bibinfo {pages} {55006} (\bibinfo {year} {2009})}\BibitemShut {NoStop}%
\bibitem [{\citenamefont {{Isayama}}, \citenamefont {{Matsunaga}},\ and\
  \citenamefont {{Hirano}}(2013)}]{IsayamaJPFR13}%
  \BibitemOpen
  \bibfield  {author} {\bibinfo {author} {\bibfnamefont {A.}~\bibnamefont
  {{Isayama}}}, \bibinfo {author} {\bibfnamefont {G.}~\bibnamefont
  {{Matsunaga}}},\ and\ \bibinfo {author} {\bibfnamefont {Y.}~\bibnamefont
  {{Hirano}}},\ }\bibfield  {title} {\enquote {\bibinfo {title} {{Onset and
  Evolution of m/n = 2/1 Neoclassical Tearing Modes in
  High-{\ensuremath{\beta}}p Mode Discharges in JT-60U}},}\ }\href
  {https://doi.org/10.1585/pfr.8.1402013} {\bibfield  {journal} {\bibinfo
  {journal} {Plasma and Fusion Research}\ }\textbf {\bibinfo {volume} {8}},\
  \bibinfo {eid} {1402013-1402013} (\bibinfo {year} {2013})}\BibitemShut
  {NoStop}%
\bibitem [{\citenamefont {Poli}\ \emph {et~al.}(2016)\citenamefont {Poli},
  \citenamefont {Bergmann}, \citenamefont {Casson}, \citenamefont {Hornsby},
  \citenamefont {Peeters}, \citenamefont {Siccinio},\ and\ \citenamefont
  {Zarzoso}}]{PoliPPR16}%
  \BibitemOpen
  \bibfield  {author} {\bibinfo {author} {\bibfnamefont {E.}~\bibnamefont
  {Poli}}, \bibinfo {author} {\bibfnamefont {A.}~\bibnamefont {Bergmann}},
  \bibinfo {author} {\bibfnamefont {F.~J.}\ \bibnamefont {Casson}}, \bibinfo
  {author} {\bibfnamefont {W.~A.}\ \bibnamefont {Hornsby}}, \bibinfo {author}
  {\bibfnamefont {A.~G.}\ \bibnamefont {Peeters}}, \bibinfo {author}
  {\bibfnamefont {M.}~\bibnamefont {Siccinio}},\ and\ \bibinfo {author}
  {\bibfnamefont {D.}~\bibnamefont {Zarzoso}},\ }\bibfield  {title} {\enquote
  {\bibinfo {title} {{Kinetic Effects on the Currents Determining the Stability
  of a Magnetic Island in Tokamaks}},}\ }\href
  {https://doi.org/10.1134/S1063780X16050135} {\bibfield  {journal} {\bibinfo
  {journal} {Plasma Physics Reports}\ }\textbf {\bibinfo {volume} {42}},\
  \bibinfo {pages} {450--464} (\bibinfo {year} {2016})}\BibitemShut {NoStop}%
\bibitem [{\citenamefont {{Ishizawa}}, \citenamefont {{Kishimoto}},\ and\
  \citenamefont {{Nakamura}}(2019)}]{IshizawaPPCF19}%
  \BibitemOpen
  \bibfield  {author} {\bibinfo {author} {\bibfnamefont {A.}~\bibnamefont
  {{Ishizawa}}}, \bibinfo {author} {\bibfnamefont {Y.}~\bibnamefont
  {{Kishimoto}}},\ and\ \bibinfo {author} {\bibfnamefont {Y.}~\bibnamefont
  {{Nakamura}}},\ }\bibfield  {title} {\enquote {\bibinfo {title} {{Multi-scale
  interactions between turbulence and magnetic islands and parity
  mixture{\textemdash}a review}},}\ }\href
  {https://doi.org/10.1088/1361-6587/ab06a8} {\bibfield  {journal} {\bibinfo
  {journal} {Plasma Physics and Controlled Fusion}\ }\textbf {\bibinfo {volume}
  {61}},\ \bibinfo {eid} {054006} (\bibinfo {year} {2019})}\BibitemShut
  {NoStop}%
\bibitem [{\citenamefont {Choi}(2021)}]{Choi2021a}%
  \BibitemOpen
  \bibfield  {author} {\bibinfo {author} {\bibfnamefont {M.~J.}\ \bibnamefont
  {Choi}},\ }\href {https://doi.org/10.1007/s41614-021-00058-w} {\emph
  {\bibinfo {title} {Reviews of Modern Plasma Physics}}},\ Vol.~\bibinfo
  {volume} {5}\ (\bibinfo  {publisher} {Springer Singapore},\ \bibinfo {year}
  {2021})\ pp.\ \bibinfo {pages} {1--39}\BibitemShut {NoStop}%
\bibitem [{\citenamefont {Furth}, \citenamefont {Rutherford},\ and\
  \citenamefont {Selberg}(1973)}]{furth_PF1973}%
  \BibitemOpen
  \bibfield  {author} {\bibinfo {author} {\bibfnamefont {H.~P.}\ \bibnamefont
  {Furth}}, \bibinfo {author} {\bibfnamefont {P.~H.}\ \bibnamefont
  {Rutherford}},\ and\ \bibinfo {author} {\bibfnamefont {H.}~\bibnamefont
  {Selberg}},\ }\bibfield  {title} {\enquote {\bibinfo {title} {{Tearing mode
  in the cylindrical tokamak}},}\ }\href {https://doi.org/10.1063/1.1694467}
  {\bibfield  {journal} {\bibinfo  {journal} {Physics of Fluids}\ }\textbf
  {\bibinfo {volume} {16}},\ \bibinfo {pages} {1054--1063} (\bibinfo {year}
  {1973})}\BibitemShut {NoStop}%
\bibitem [{\citenamefont {{Muraglia}}\ \emph {et~al.}(2011)\citenamefont
  {{Muraglia}}, \citenamefont {{Agullo}}, \citenamefont {{Benkadda}},
  \citenamefont {{Yagi}}, \citenamefont {{Garbet}},\ and\ \citenamefont
  {{Sen}}}]{MuragliaPRL11}%
  \BibitemOpen
  \bibfield  {author} {\bibinfo {author} {\bibfnamefont {M.}~\bibnamefont
  {{Muraglia}}}, \bibinfo {author} {\bibfnamefont {O.}~\bibnamefont
  {{Agullo}}}, \bibinfo {author} {\bibfnamefont {S.}~\bibnamefont
  {{Benkadda}}}, \bibinfo {author} {\bibfnamefont {M.}~\bibnamefont {{Yagi}}},
  \bibinfo {author} {\bibfnamefont {X.}~\bibnamefont {{Garbet}}},\ and\
  \bibinfo {author} {\bibfnamefont {A.}~\bibnamefont {{Sen}}},\ }\bibfield
  {title} {\enquote {\bibinfo {title} {{Generation and Amplification of
  Magnetic Islands by Drift Interchange Turbulence}},}\ }\href
  {https://doi.org/10.1103/PhysRevLett.107.095003} {\bibfield  {journal}
  {\bibinfo  {journal} {\prl}\ }\textbf {\bibinfo {volume} {107}},\ \bibinfo
  {eid} {095003} (\bibinfo {year} {2011})}\BibitemShut {NoStop}%
\bibitem [{\citenamefont {Ishizawa}\ \emph {et~al.}(2015)\citenamefont
  {Ishizawa}, \citenamefont {Maeyama}, \citenamefont {Watanabe}, \citenamefont
  {Sugama},\ and\ \citenamefont {Nakajima}}]{IshizawaJPP15}%
  \BibitemOpen
  \bibfield  {author} {\bibinfo {author} {\bibfnamefont {A.}~\bibnamefont
  {Ishizawa}}, \bibinfo {author} {\bibfnamefont {S.}~\bibnamefont {Maeyama}},
  \bibinfo {author} {\bibfnamefont {T.~H.}\ \bibnamefont {Watanabe}}, \bibinfo
  {author} {\bibfnamefont {H.}~\bibnamefont {Sugama}},\ and\ \bibinfo {author}
  {\bibfnamefont {N.}~\bibnamefont {Nakajima}},\ }\bibfield  {title} {\enquote
  {\bibinfo {title} {{Electromagnetic gyrokinetic simulation of turbulence in
  torus plasmas}},}\ }\href {https://doi.org/10.1017/S0022377815000100}
  {\bibfield  {journal} {\bibinfo  {journal} {Journal of Plasma Physics}\
  }\textbf {\bibinfo {volume} {81}} (\bibinfo {year} {2015}),\
  10.1017/S0022377815000100}\BibitemShut {NoStop}%
\bibitem [{\citenamefont {{Fietz}}\ \emph {et~al.}(2013)\citenamefont
  {{Fietz}}, \citenamefont {{Maraschek}}, \citenamefont {{Zohm}}, \citenamefont
  {{Reich}}, \citenamefont {{Barrera}}, \citenamefont {{McDermott}},\ and\
  \citenamefont {{the ASDEX Upgrade Team}}}]{FietzPPCF13}%
  \BibitemOpen
  \bibfield  {author} {\bibinfo {author} {\bibfnamefont {S.}~\bibnamefont
  {{Fietz}}}, \bibinfo {author} {\bibfnamefont {M.}~\bibnamefont
  {{Maraschek}}}, \bibinfo {author} {\bibfnamefont {H.}~\bibnamefont {{Zohm}}},
  \bibinfo {author} {\bibfnamefont {M.}~\bibnamefont {{Reich}}}, \bibinfo
  {author} {\bibfnamefont {L.}~\bibnamefont {{Barrera}}}, \bibinfo {author}
  {\bibfnamefont {R.~M.}\ \bibnamefont {{McDermott}}},\ and\ \bibinfo {author}
  {\bibnamefont {{the ASDEX Upgrade Team}}},\ }\bibfield  {title} {\enquote
  {\bibinfo {title} {{Influence of rotation on the (m, n) = (3, 2) neoclassical
  tearing mode threshold in the ASDEX Upgrade}},}\ }\href
  {https://doi.org/10.1088/0741-3335/55/8/085010} {\bibfield  {journal}
  {\bibinfo  {journal} {Plasma Physics and Controlled Fusion}\ }\textbf
  {\bibinfo {volume} {55}},\ \bibinfo {eid} {085010} (\bibinfo {year}
  {2013})}\BibitemShut {NoStop}%
\bibitem [{\citenamefont {{Brizard}}\ and\ \citenamefont
  {{Hahm}}(2007)}]{BrizardRMP07}%
  \BibitemOpen
  \bibfield  {author} {\bibinfo {author} {\bibfnamefont {A.~J.}\ \bibnamefont
  {{Brizard}}}\ and\ \bibinfo {author} {\bibfnamefont {T.~S.}\ \bibnamefont
  {{Hahm}}},\ }\bibfield  {title} {\enquote {\bibinfo {title} {{Foundations of
  nonlinear gyrokinetic theory}},}\ }\href
  {https://doi.org/10.1103/RevModPhys.79.421} {\bibfield  {journal} {\bibinfo
  {journal} {Reviews of Modern Physics}\ }\textbf {\bibinfo {volume} {79}},\
  \bibinfo {pages} {421--468} (\bibinfo {year} {2007})}\BibitemShut {NoStop}%
\bibitem [{\citenamefont {{Poli}}, \citenamefont {{Bottino}},\ and\
  \citenamefont {{Peeters}}(2009)}]{PoliNF09}%
  \BibitemOpen
  \bibfield  {author} {\bibinfo {author} {\bibfnamefont {E.}~\bibnamefont
  {{Poli}}}, \bibinfo {author} {\bibfnamefont {A.}~\bibnamefont {{Bottino}}},\
  and\ \bibinfo {author} {\bibfnamefont {A.~G.}\ \bibnamefont {{Peeters}}},\
  }\bibfield  {title} {\enquote {\bibinfo {title} {{Behaviour of turbulent
  transport in the vicinity of a magnetic island}},}\ }\href
  {https://doi.org/10.1088/0029-5515/49/7/075010} {\bibfield  {journal}
  {\bibinfo  {journal} {Nuclear Fusion}\ }\textbf {\bibinfo {volume} {49}},\
  \bibinfo {eid} {075010} (\bibinfo {year} {2009})}\BibitemShut {NoStop}%
\bibitem [{\citenamefont {Poli}\ \emph {et~al.}(2010)\citenamefont {Poli},
  \citenamefont {Bottino}, \citenamefont {Hornsby}, \citenamefont {Peeters},
  \citenamefont {Ribeiro}, \citenamefont {Scott},\ and\ \citenamefont
  {Siccinio}}]{PoliPPCF10a}%
  \BibitemOpen
  \bibfield  {author} {\bibinfo {author} {\bibfnamefont {E.}~\bibnamefont
  {Poli}}, \bibinfo {author} {\bibfnamefont {A.}~\bibnamefont {Bottino}},
  \bibinfo {author} {\bibfnamefont {W.~A.}\ \bibnamefont {Hornsby}}, \bibinfo
  {author} {\bibfnamefont {A.~G.}\ \bibnamefont {Peeters}}, \bibinfo {author}
  {\bibfnamefont {T.}~\bibnamefont {Ribeiro}}, \bibinfo {author} {\bibfnamefont
  {B.~D.}\ \bibnamefont {Scott}},\ and\ \bibinfo {author} {\bibfnamefont
  {M.}~\bibnamefont {Siccinio}},\ }\bibfield  {title} {\enquote {\bibinfo
  {title} {{Gyrokinetic and gyrofluid investigation of magnetic islands in
  tokamaks}},}\ }\href {https://doi.org/10.1088/0741-3335/52/12/124021}
  {\bibfield  {journal} {\bibinfo  {journal} {Plasma Physics and Controlled
  Fusion}\ }\textbf {\bibinfo {volume} {52}},\ \bibinfo {pages} {124021}
  (\bibinfo {year} {2010})}\BibitemShut {NoStop}%
\bibitem [{\citenamefont {{Jiang}}\ \emph {et~al.}(2014)\citenamefont
  {{Jiang}}, \citenamefont {{Lin}}, \citenamefont {{Holod}},\ and\
  \citenamefont {{Xiao}}}]{JiangPoP14}%
  \BibitemOpen
  \bibfield  {author} {\bibinfo {author} {\bibfnamefont {P.}~\bibnamefont
  {{Jiang}}}, \bibinfo {author} {\bibfnamefont {Z.}~\bibnamefont {{Lin}}},
  \bibinfo {author} {\bibfnamefont {I.}~\bibnamefont {{Holod}}},\ and\ \bibinfo
  {author} {\bibfnamefont {C.}~\bibnamefont {{Xiao}}},\ }\bibfield  {title}
  {\enquote {\bibinfo {title} {{Effects of magnetic islands on drift wave
  instability}},}\ }\href {https://doi.org/10.1063/1.4903910} {\bibfield
  {journal} {\bibinfo  {journal} {Physics of Plasmas}\ }\textbf {\bibinfo
  {volume} {21}},\ \bibinfo {eid} {122513} (\bibinfo {year}
  {2014})}\BibitemShut {NoStop}%
\bibitem [{\citenamefont {{Zarzoso}}\ \emph {et~al.}(2015)\citenamefont
  {{Zarzoso}}, \citenamefont {{Hornsby}}, \citenamefont {{Poli}}, \citenamefont
  {{Casson}}, \citenamefont {{Peeters}},\ and\ \citenamefont
  {{Nasr}}}]{ZarzosoNF15}%
  \BibitemOpen
  \bibfield  {author} {\bibinfo {author} {\bibfnamefont {D.}~\bibnamefont
  {{Zarzoso}}}, \bibinfo {author} {\bibfnamefont {W.~A.}\ \bibnamefont
  {{Hornsby}}}, \bibinfo {author} {\bibfnamefont {E.}~\bibnamefont {{Poli}}},
  \bibinfo {author} {\bibfnamefont {F.~J.}\ \bibnamefont {{Casson}}}, \bibinfo
  {author} {\bibfnamefont {A.~G.}\ \bibnamefont {{Peeters}}},\ and\ \bibinfo
  {author} {\bibfnamefont {S.}~\bibnamefont {{Nasr}}},\ }\bibfield  {title}
  {\enquote {\bibinfo {title} {{Impact of rotating magnetic islands on density
  profile flattening and turbulent transport}},}\ }\href
  {https://doi.org/10.1088/0029-5515/55/11/113018} {\bibfield  {journal}
  {\bibinfo  {journal} {Nuclear Fusion}\ }\textbf {\bibinfo {volume} {55}},\
  \bibinfo {eid} {113018} (\bibinfo {year} {2015})}\BibitemShut {NoStop}%
\bibitem [{\citenamefont {{Muto}}, \citenamefont {{Imadera}},\ and\
  \citenamefont {{Kishimoto}}(2022)}]{MutoPoP22}%
  \BibitemOpen
  \bibfield  {author} {\bibinfo {author} {\bibfnamefont {M.}~\bibnamefont
  {{Muto}}}, \bibinfo {author} {\bibfnamefont {K.}~\bibnamefont {{Imadera}}},\
  and\ \bibinfo {author} {\bibfnamefont {Y.}~\bibnamefont {{Kishimoto}}},\
  }\bibfield  {title} {\enquote {\bibinfo {title} {{Effects of magnetic island
  on profile formation in flux-driven ITG turbulence}},}\ }\href
  {https://doi.org/10.1063/5.0081125} {\bibfield  {journal} {\bibinfo
  {journal} {Physics of Plasmas}\ }\textbf {\bibinfo {volume} {29}},\ \bibinfo
  {eid} {052503} (\bibinfo {year} {2022})}\BibitemShut {NoStop}%
\bibitem [{\citenamefont {{Li}}\ \emph {et~al.}(2023)\citenamefont {{Li}},
  \citenamefont {{Xu}}, \citenamefont {{Qu}}, \citenamefont {{Lin}},
  \citenamefont {{Dong}}, \citenamefont {{Peng}},\ and\ \citenamefont
  {{Li}}}]{LiNF23}%
  \BibitemOpen
  \bibfield  {author} {\bibinfo {author} {\bibfnamefont {J.}~\bibnamefont
  {{Li}}}, \bibinfo {author} {\bibfnamefont {J.~Q.}\ \bibnamefont {{Xu}}},
  \bibinfo {author} {\bibfnamefont {Y.~R.}\ \bibnamefont {{Qu}}}, \bibinfo
  {author} {\bibfnamefont {Z.}~\bibnamefont {{Lin}}}, \bibinfo {author}
  {\bibfnamefont {J.~Q.}\ \bibnamefont {{Dong}}}, \bibinfo {author}
  {\bibfnamefont {X.~D.}\ \bibnamefont {{Peng}}},\ and\ \bibinfo {author}
  {\bibfnamefont {J.~Q.}\ \bibnamefont {{Li}}},\ }\bibfield  {title} {\enquote
  {\bibinfo {title} {{Global gyrokinetic simulations of the impact of magnetic
  island on ion temperature gradient driven turbulence}},}\ }\href
  {https://doi.org/10.1088/1741-4326/ace461} {\bibfield  {journal} {\bibinfo
  {journal} {Nuclear Fusion}\ }\textbf {\bibinfo {volume} {63}},\ \bibinfo
  {eid} {096005} (\bibinfo {year} {2023})},\ \Eprint
  {https://arxiv.org/abs/2306.05607} {arXiv:2306.05607 [physics.plasm-ph]}
  \BibitemShut {NoStop}%
\bibitem [{\citenamefont {Hornsby}\ \emph {et~al.}(2015)\citenamefont
  {Hornsby}, \citenamefont {Migliano}, \citenamefont {Buchholz}, \citenamefont
  {Grosshauser}, \citenamefont {Weikl}, \citenamefont {Zarzoso}, \citenamefont
  {Casson}, \citenamefont {Poli},\ and\ \citenamefont
  {Peeters}}]{HornsbyPPCF15}%
  \BibitemOpen
  \bibfield  {author} {\bibinfo {author} {\bibfnamefont {W.~A.}\ \bibnamefont
  {Hornsby}}, \bibinfo {author} {\bibfnamefont {P.}~\bibnamefont {Migliano}},
  \bibinfo {author} {\bibfnamefont {R.}~\bibnamefont {Buchholz}}, \bibinfo
  {author} {\bibfnamefont {S.}~\bibnamefont {Grosshauser}}, \bibinfo {author}
  {\bibfnamefont {A.}~\bibnamefont {Weikl}}, \bibinfo {author} {\bibfnamefont
  {D.}~\bibnamefont {Zarzoso}}, \bibinfo {author} {\bibfnamefont {F.~J.}\
  \bibnamefont {Casson}}, \bibinfo {author} {\bibfnamefont {E.}~\bibnamefont
  {Poli}},\ and\ \bibinfo {author} {\bibfnamefont {A.~G.}\ \bibnamefont
  {Peeters}},\ }\bibfield  {title} {\enquote {\bibinfo {title} {On seed island
  generation and the non-linear self-consistent interaction of the tearing mode
  with electromagnetic gyro-kinetic turbulence},}\ }\href
  {https://doi.org/10.1088/0741-3335/57/5/054018} {\bibfield  {journal}
  {\bibinfo  {journal} {Plasma Physics and Controlled Fusion}\ }\textbf
  {\bibinfo {volume} {57}},\ \bibinfo {pages} {054018} (\bibinfo {year}
  {2015})}\BibitemShut {NoStop}%
\bibitem [{\citenamefont {Hornsby}\ \emph {et~al.}(2016)\citenamefont
  {Hornsby}, \citenamefont {Migliano}, \citenamefont {Buchholz}, \citenamefont
  {Grosshauser}, \citenamefont {Weikl}, \citenamefont {Zarzoso}, \citenamefont
  {Casson}, \citenamefont {Poli},\ and\ \citenamefont
  {Peeters}}]{HornsbyPPCF16}%
  \BibitemOpen
  \bibfield  {author} {\bibinfo {author} {\bibfnamefont {W.~A.}\ \bibnamefont
  {Hornsby}}, \bibinfo {author} {\bibfnamefont {P.}~\bibnamefont {Migliano}},
  \bibinfo {author} {\bibfnamefont {R.}~\bibnamefont {Buchholz}}, \bibinfo
  {author} {\bibfnamefont {S.}~\bibnamefont {Grosshauser}}, \bibinfo {author}
  {\bibfnamefont {A.}~\bibnamefont {Weikl}}, \bibinfo {author} {\bibfnamefont
  {D.}~\bibnamefont {Zarzoso}}, \bibinfo {author} {\bibfnamefont {F.~J.}\
  \bibnamefont {Casson}}, \bibinfo {author} {\bibfnamefont {E.}~\bibnamefont
  {Poli}},\ and\ \bibinfo {author} {\bibfnamefont {A.~G.}\ \bibnamefont
  {Peeters}},\ }\bibfield  {title} {\enquote {\bibinfo {title} {{The non-linear
  evolution of the tearing mode in electromagnetic turbulence using gyrokinetic
  simulations}},}\ }\href {https://doi.org/10.1088/0741-3335/58/1/014028}
  {\bibfield  {journal} {\bibinfo  {journal} {Plasma Physics and Controlled
  Fusion}\ }\textbf {\bibinfo {volume} {58}},\ \bibinfo {pages} {014028}
  (\bibinfo {year} {2016})}\BibitemShut {NoStop}%
\bibitem [{\citenamefont {{Jitsuk}}\ \emph {et~al.}(2024)\citenamefont
  {{Jitsuk}}, \citenamefont {{Di Siena}}, \citenamefont {{Pueschel}},
  \citenamefont {{Terry}}, \citenamefont {{Widmer}}, \citenamefont {{Poli}},\
  and\ \citenamefont {{Sarff}}}]{JitsukNF24}%
  \BibitemOpen
  \bibfield  {author} {\bibinfo {author} {\bibfnamefont {T.}~\bibnamefont
  {{Jitsuk}}}, \bibinfo {author} {\bibfnamefont {A.}~\bibnamefont {{Di
  Siena}}}, \bibinfo {author} {\bibfnamefont {M.~J.}\ \bibnamefont
  {{Pueschel}}}, \bibinfo {author} {\bibfnamefont {P.~W.}\ \bibnamefont
  {{Terry}}}, \bibinfo {author} {\bibfnamefont {F.}~\bibnamefont {{Widmer}}},
  \bibinfo {author} {\bibfnamefont {E.}~\bibnamefont {{Poli}}},\ and\ \bibinfo
  {author} {\bibfnamefont {J.~S.}\ \bibnamefont {{Sarff}}},\ }\bibfield
  {title} {\enquote {\bibinfo {title} {{Global linear and nonlinear gyrokinetic
  simulations of tearing modes}},}\ }\href
  {https://doi.org/10.1088/1741-4326/ad279b} {\bibfield  {journal} {\bibinfo
  {journal} {Nuclear Fusion}\ }\textbf {\bibinfo {volume} {64}},\ \bibinfo
  {eid} {046005} (\bibinfo {year} {2024})},\ \Eprint
  {https://arxiv.org/abs/2308.16345} {arXiv:2308.16345 [physics.plasm-ph]}
  \BibitemShut {NoStop}%
\bibitem [{\citenamefont {{Widmer}}\ \emph {et~al.}(2024)\citenamefont
  {{Widmer}}, \citenamefont {{Poli}}, \citenamefont {{Mishchenko}},
  \citenamefont {{Ishizawa}}, \citenamefont {{Bottino}},\ and\ \citenamefont
  {{Hayward-Schneider}}}]{WidmerPoP24}%
  \BibitemOpen
  \bibfield  {author} {\bibinfo {author} {\bibfnamefont {F.}~\bibnamefont
  {{Widmer}}}, \bibinfo {author} {\bibfnamefont {E.}~\bibnamefont {{Poli}}},
  \bibinfo {author} {\bibfnamefont {A.}~\bibnamefont {{Mishchenko}}}, \bibinfo
  {author} {\bibfnamefont {A.}~\bibnamefont {{Ishizawa}}}, \bibinfo {author}
  {\bibfnamefont {A.}~\bibnamefont {{Bottino}}},\ and\ \bibinfo {author}
  {\bibfnamefont {T.}~\bibnamefont {{Hayward-Schneider}}},\ }\bibfield  {title}
  {\enquote {\bibinfo {title} {{Linear and nonlinear dynamics of
  self-consistent collisionless tearing modes in toroidal gyrokinetic
  simulations}},}\ }\href {https://doi.org/10.1063/5.0221751} {\bibfield
  {journal} {\bibinfo  {journal} {Physics of Plasmas}\ }\textbf {\bibinfo
  {volume} {31}},\ \bibinfo {eid} {112505} (\bibinfo {year} {2024})},\ \Eprint
  {https://arxiv.org/abs/2410.11498} {arXiv:2410.11498 [physics.plasm-ph]}
  \BibitemShut {NoStop}%
\bibitem [{\citenamefont {{Wei}}\ \emph {et~al.}(2025)\citenamefont {{Wei}},
  \citenamefont {{Nicolau}}, \citenamefont {{Choi}}, \citenamefont {{Lin}},
  \citenamefont {{Yang}}, \citenamefont {{Kim}}, \citenamefont {{Lee}},
  \citenamefont {{Zhao}}, \citenamefont {{Cote}}, \citenamefont {{Park}},\ and\
  \citenamefont {{Orlov}}}]{WeiNF25}%
  \BibitemOpen
  \bibfield  {author} {\bibinfo {author} {\bibfnamefont {X.}~\bibnamefont
  {{Wei}}}, \bibinfo {author} {\bibfnamefont {J.~H.}\ \bibnamefont
  {{Nicolau}}}, \bibinfo {author} {\bibfnamefont {G.}~\bibnamefont {{Choi}}},
  \bibinfo {author} {\bibfnamefont {Z.}~\bibnamefont {{Lin}}}, \bibinfo
  {author} {\bibfnamefont {S.-M.}\ \bibnamefont {{Yang}}}, \bibinfo {author}
  {\bibfnamefont {S.}~\bibnamefont {{Kim}}}, \bibinfo {author} {\bibfnamefont
  {W.}~\bibnamefont {{Lee}}}, \bibinfo {author} {\bibfnamefont
  {C.}~\bibnamefont {{Zhao}}}, \bibinfo {author} {\bibfnamefont
  {T.}~\bibnamefont {{Cote}}}, \bibinfo {author} {\bibfnamefont
  {J.}~\bibnamefont {{Park}}},\ and\ \bibinfo {author} {\bibfnamefont
  {D.}~\bibnamefont {{Orlov}}},\ }\bibfield  {title} {\enquote {\bibinfo
  {title} {{Gyrokinetic simulations of the effects of magnetic islands on
  microturbulence in KSTAR}},}\ }\href
  {https://doi.org/10.1088/1741-4326/ada049} {\bibfield  {journal} {\bibinfo
  {journal} {Nuclear Fusion}\ }\textbf {\bibinfo {volume} {65}},\ \bibinfo
  {eid} {026026} (\bibinfo {year} {2025})},\ \Eprint
  {https://arxiv.org/abs/2412.09522} {arXiv:2412.09522 [physics.plasm-ph]}
  \BibitemShut {NoStop}%
\bibitem [{\citenamefont {{Lanti}}\ \emph {et~al.}(2020)\citenamefont
  {{Lanti}}, \citenamefont {{Ohana}}, \citenamefont {{Tronko}}, \citenamefont
  {{Hayward-Schneider}}, \citenamefont {{Bottino}}, \citenamefont {{McMillan}},
  \citenamefont {{Mishchenko}}, \citenamefont {{Scheinberg}}, \citenamefont
  {{Biancalani}}, \citenamefont {{Angelino}}, \citenamefont {{Brunner}},
  \citenamefont {{Dominski}}, \citenamefont {{Donnel}}, \citenamefont
  {{Gheller}}, \citenamefont {{Hatzky}}, \citenamefont {{Jocksch}},
  \citenamefont {{Jolliet}}, \citenamefont {{Lu}}, \citenamefont {{Martin
  Collar}}, \citenamefont {{Novikau}}, \citenamefont {{Sonnendr{\"u}cker}},
  \citenamefont {{Vernay}},\ and\ \citenamefont {{Villard}}}]{LantiCPC20}%
  \BibitemOpen
  \bibfield  {author} {\bibinfo {author} {\bibfnamefont {E.}~\bibnamefont
  {{Lanti}}}, \bibinfo {author} {\bibfnamefont {N.}~\bibnamefont {{Ohana}}},
  \bibinfo {author} {\bibfnamefont {N.}~\bibnamefont {{Tronko}}}, \bibinfo
  {author} {\bibfnamefont {T.}~\bibnamefont {{Hayward-Schneider}}}, \bibinfo
  {author} {\bibfnamefont {A.}~\bibnamefont {{Bottino}}}, \bibinfo {author}
  {\bibfnamefont {B.~F.}\ \bibnamefont {{McMillan}}}, \bibinfo {author}
  {\bibfnamefont {A.}~\bibnamefont {{Mishchenko}}}, \bibinfo {author}
  {\bibfnamefont {A.}~\bibnamefont {{Scheinberg}}}, \bibinfo {author}
  {\bibfnamefont {A.}~\bibnamefont {{Biancalani}}}, \bibinfo {author}
  {\bibfnamefont {P.}~\bibnamefont {{Angelino}}}, \bibinfo {author}
  {\bibfnamefont {S.}~\bibnamefont {{Brunner}}}, \bibinfo {author}
  {\bibfnamefont {J.}~\bibnamefont {{Dominski}}}, \bibinfo {author}
  {\bibfnamefont {P.}~\bibnamefont {{Donnel}}}, \bibinfo {author}
  {\bibfnamefont {C.}~\bibnamefont {{Gheller}}}, \bibinfo {author}
  {\bibfnamefont {R.}~\bibnamefont {{Hatzky}}}, \bibinfo {author}
  {\bibfnamefont {A.}~\bibnamefont {{Jocksch}}}, \bibinfo {author}
  {\bibfnamefont {S.}~\bibnamefont {{Jolliet}}}, \bibinfo {author}
  {\bibfnamefont {Z.~X.}\ \bibnamefont {{Lu}}}, \bibinfo {author}
  {\bibfnamefont {J.~P.}\ \bibnamefont {{Martin Collar}}}, \bibinfo {author}
  {\bibfnamefont {I.}~\bibnamefont {{Novikau}}}, \bibinfo {author}
  {\bibfnamefont {E.}~\bibnamefont {{Sonnendr{\"u}cker}}}, \bibinfo {author}
  {\bibfnamefont {T.}~\bibnamefont {{Vernay}}},\ and\ \bibinfo {author}
  {\bibfnamefont {L.}~\bibnamefont {{Villard}}},\ }\bibfield  {title} {\enquote
  {\bibinfo {title} {{ORB5: A global electromagnetic gyrokinetic code using the
  PIC approach in toroidal geometry}},}\ }\href
  {https://doi.org/10.1016/j.cpc.2019.107072} {\bibfield  {journal} {\bibinfo
  {journal} {Computer Physics Communications}\ }\textbf {\bibinfo {volume}
  {251}},\ \bibinfo {eid} {107072} (\bibinfo {year} {2020})},\ \Eprint
  {https://arxiv.org/abs/1905.01906} {arXiv:1905.01906 [physics.plasm-ph]}
  \BibitemShut {NoStop}%
\bibitem [{\citenamefont {Mishchenko}\ \emph {et~al.}(2019)\citenamefont
  {Mishchenko}, \citenamefont {Bottino}, \citenamefont {Biancalani},
  \citenamefont {Hatzky}, \citenamefont {Hayward-Schneider}, \citenamefont
  {Ohana}, \citenamefont {Lanti}, \citenamefont {Brunner}, \citenamefont
  {Villard}, \citenamefont {Borchardt}, \citenamefont {Kleiber},\ and\
  \citenamefont {K{\"{o}}nies}}]{MishchenkoCPC19}%
  \BibitemOpen
  \bibfield  {author} {\bibinfo {author} {\bibfnamefont {A.}~\bibnamefont
  {Mishchenko}}, \bibinfo {author} {\bibfnamefont {A.}~\bibnamefont {Bottino}},
  \bibinfo {author} {\bibfnamefont {A.}~\bibnamefont {Biancalani}}, \bibinfo
  {author} {\bibfnamefont {R.}~\bibnamefont {Hatzky}}, \bibinfo {author}
  {\bibfnamefont {T.}~\bibnamefont {Hayward-Schneider}}, \bibinfo {author}
  {\bibfnamefont {N.}~\bibnamefont {Ohana}}, \bibinfo {author} {\bibfnamefont
  {E.}~\bibnamefont {Lanti}}, \bibinfo {author} {\bibfnamefont
  {S.}~\bibnamefont {Brunner}}, \bibinfo {author} {\bibfnamefont
  {L.}~\bibnamefont {Villard}}, \bibinfo {author} {\bibfnamefont
  {M.}~\bibnamefont {Borchardt}}, \bibinfo {author} {\bibfnamefont
  {R.}~\bibnamefont {Kleiber}},\ and\ \bibinfo {author} {\bibfnamefont
  {A.}~\bibnamefont {K{\"{o}}nies}},\ }\bibfield  {title} {\enquote {\bibinfo
  {title} {{Pullback scheme implementation in ORB5}},}\ }\href
  {https://doi.org/10.1016/j.cpc.2018.12.002} {\bibfield  {journal} {\bibinfo
  {journal} {Computer Physics Communications}\ }\textbf {\bibinfo {volume}
  {238}},\ \bibinfo {pages} {194--202} (\bibinfo {year} {2019})},\ \Eprint
  {https://arxiv.org/abs/1811.05346} {arXiv:1811.05346} \BibitemShut {NoStop}%
\bibitem [{\citenamefont {{Drake}}\ and\ \citenamefont
  {{Lee}}(1977)}]{Drake77}%
  \BibitemOpen
  \bibfield  {author} {\bibinfo {author} {\bibfnamefont {J.~F.}\ \bibnamefont
  {{Drake}}}\ and\ \bibinfo {author} {\bibfnamefont {Y.~C.}\ \bibnamefont
  {{Lee}}},\ }\bibfield  {title} {\enquote {\bibinfo {title} {{Kinetic theory
  of tearing instabilities}},}\ }\href {https://doi.org/10.1063/1.862017}
  {\bibfield  {journal} {\bibinfo  {journal} {Physics of Fluids}\ }\textbf
  {\bibinfo {volume} {20}},\ \bibinfo {pages} {1341--1353} (\bibinfo {year}
  {1977})}\BibitemShut {NoStop}%
\bibitem [{\citenamefont {Hamed}\ \emph {et~al.}(2019)\citenamefont {Hamed},
  \citenamefont {Muraglia}, \citenamefont {Camenen}, \citenamefont {Garbet},\
  and\ \citenamefont {Agullo}}]{HamedPoP19}%
  \BibitemOpen
  \bibfield  {author} {\bibinfo {author} {\bibfnamefont {M.}~\bibnamefont
  {Hamed}}, \bibinfo {author} {\bibfnamefont {M.}~\bibnamefont {Muraglia}},
  \bibinfo {author} {\bibfnamefont {Y.}~\bibnamefont {Camenen}}, \bibinfo
  {author} {\bibfnamefont {X.}~\bibnamefont {Garbet}},\ and\ \bibinfo {author}
  {\bibfnamefont {O.}~\bibnamefont {Agullo}},\ }\bibfield  {title} {\enquote
  {\bibinfo {title} {{Impact of electric potential and magnetic drift on
  microtearing modes stability}},}\ }\href {https://doi.org/10.1063/1.5111701}
  {\bibfield  {journal} {\bibinfo  {journal} {Physics of Plasmas}\ }\textbf
  {\bibinfo {volume} {26}} (\bibinfo {year} {2019}),\
  10.1063/1.5111701}\BibitemShut {NoStop}%
\bibitem [{\citenamefont {Xie}, \citenamefont {Mahajan},\ and\ \citenamefont
  {Hatch}(2023)}]{XiePoP23}%
  \BibitemOpen
  \bibfield  {author} {\bibinfo {author} {\bibfnamefont {T.}~\bibnamefont
  {Xie}}, \bibinfo {author} {\bibfnamefont {S.~M.}\ \bibnamefont {Mahajan}},\
  and\ \bibinfo {author} {\bibfnamefont {D.~R.}\ \bibnamefont {Hatch}},\
  }\bibfield  {title} {\enquote {\bibinfo {title} {Ballooning theory for
  micro-tearing mode in tokamak},}\ }\href {https://doi.org/10.1063/5.0157408}
  {\bibfield  {journal} {\bibinfo  {journal} {Physics of Plasmas}\ }\textbf
  {\bibinfo {volume} {30}},\ \bibinfo {pages} {082111} (\bibinfo {year}
  {2023})}\BibitemShut {NoStop}%
\bibitem [{\citenamefont {Ryter}\ \emph {et~al.}(2005)\citenamefont {Ryter},
  \citenamefont {Angioni}, \citenamefont {Peeters}, \citenamefont {Leuterer},
  \citenamefont {Fahrbach},\ and\ \citenamefont {Suttrop}}]{RyterPRL05}%
  \BibitemOpen
  \bibfield  {author} {\bibinfo {author} {\bibfnamefont {F.}~\bibnamefont
  {Ryter}}, \bibinfo {author} {\bibfnamefont {C.}~\bibnamefont {Angioni}},
  \bibinfo {author} {\bibfnamefont {A.~G.}\ \bibnamefont {Peeters}}, \bibinfo
  {author} {\bibfnamefont {F.}~\bibnamefont {Leuterer}}, \bibinfo {author}
  {\bibfnamefont {H.-U.}\ \bibnamefont {Fahrbach}},\ and\ \bibinfo {author}
  {\bibfnamefont {W.}~\bibnamefont {Suttrop}} (\bibinfo {collaboration} {ASDEX
  Upgrade Team}),\ }\bibfield  {title} {\enquote {\bibinfo {title}
  {Experimental study of trapped-electron-mode properties in tokamaks:
  Threshold and stabilization by collisions},}\ }\href
  {https://doi.org/10.1103/PhysRevLett.95.085001} {\bibfield  {journal}
  {\bibinfo  {journal} {Phys. Rev. Lett.}\ }\textbf {\bibinfo {volume} {95}},\
  \bibinfo {pages} {085001} (\bibinfo {year} {2005})}\BibitemShut {NoStop}%
\bibitem [{\citenamefont {{Dannert}}\ and\ \citenamefont
  {{Jenko}}(2005)}]{DannertPoP2005}%
  \BibitemOpen
  \bibfield  {author} {\bibinfo {author} {\bibfnamefont {T.}~\bibnamefont
  {{Dannert}}}\ and\ \bibinfo {author} {\bibfnamefont {F.}~\bibnamefont
  {{Jenko}}},\ }\bibfield  {title} {\enquote {\bibinfo {title} {{Gyrokinetic
  simulation of collisionless trapped-electron mode turbulence}},}\ }\href
  {https://doi.org/10.1063/1.1947447} {\bibfield  {journal} {\bibinfo
  {journal} {Physics of Plasmas}\ }\textbf {\bibinfo {volume} {12}},\ \bibinfo
  {eid} {072309} (\bibinfo {year} {2005})}\BibitemShut {NoStop}%
\bibitem [{\citenamefont {Nakata}\ \emph {et~al.}(2017)\citenamefont {Nakata},
  \citenamefont {Nunami}, \citenamefont {Sugama},\ and\ \citenamefont
  {Watanabe}}]{Nakata2017}%
  \BibitemOpen
  \bibfield  {author} {\bibinfo {author} {\bibfnamefont {M.}~\bibnamefont
  {Nakata}}, \bibinfo {author} {\bibfnamefont {M.}~\bibnamefont {Nunami}},
  \bibinfo {author} {\bibfnamefont {H.}~\bibnamefont {Sugama}},\ and\ \bibinfo
  {author} {\bibfnamefont {T.~H.}\ \bibnamefont {Watanabe}},\ }\bibfield
  {title} {\enquote {\bibinfo {title} {{Isotope Effects on
  Trapped-Electron-Mode Driven Turbulence and Zonal Flows in Helical and
  Tokamak Plasmas}},}\ }\href {https://doi.org/10.1103/PhysRevLett.118.165002}
  {\bibfield  {journal} {\bibinfo  {journal} {Physical Review Letters}\
  }\textbf {\bibinfo {volume} {118}},\ \bibinfo {pages} {1--6} (\bibinfo {year}
  {2017})}\BibitemShut {NoStop}%
\bibitem [{\citenamefont {Maeyama}\ \emph {et~al.}(2014)\citenamefont
  {Maeyama}, \citenamefont {Ishizawa}, \citenamefont {Watanabe}, \citenamefont
  {Nakata}, \citenamefont {Miyato}, \citenamefont {Yagi},\ and\ \citenamefont
  {Idomura}}]{Maeyama_PoP2014}%
  \BibitemOpen
  \bibfield  {author} {\bibinfo {author} {\bibfnamefont {S.}~\bibnamefont
  {Maeyama}}, \bibinfo {author} {\bibfnamefont {A.}~\bibnamefont {Ishizawa}},
  \bibinfo {author} {\bibfnamefont {T.~H.}\ \bibnamefont {Watanabe}}, \bibinfo
  {author} {\bibfnamefont {M.}~\bibnamefont {Nakata}}, \bibinfo {author}
  {\bibfnamefont {N.}~\bibnamefont {Miyato}}, \bibinfo {author} {\bibfnamefont
  {M.}~\bibnamefont {Yagi}},\ and\ \bibinfo {author} {\bibfnamefont
  {Y.}~\bibnamefont {Idomura}},\ }\bibfield  {title} {\enquote {\bibinfo
  {title} {{Comparison between kinetic-ballooning-mode-driven turbulence and
  ion-temperature-gradient-driven turbulence}},}\ }\href
  {https://doi.org/10.1063/1.4873379} {\bibfield  {journal} {\bibinfo
  {journal} {Physics of Plasmas}\ }\textbf {\bibinfo {volume} {21}} (\bibinfo
  {year} {2014}),\ 10.1063/1.4873379}\BibitemShut {NoStop}%
\bibitem [{Zoh(2014)}]{ZohmBook2022}%
  \BibitemOpen
  \enquote {\bibinfo {title} {Resistive mhd stability},}\ in\ \href
  {https://doi.org/https://doi.org/10.1002/9783527677375.ch8} {\emph {\bibinfo
  {booktitle} {Magnetohydrodynamic Stability of Tokamaks}}}\ (\bibinfo
  {publisher} {John Wiley \& Sons, Ltd},\ \bibinfo {year} {2014})\
  Chap.~\bibinfo {chapter} {8}, pp.\ \bibinfo {pages} {123--140},\ \Eprint
  {https://arxiv.org/abs/https://onlinelibrary.wiley.com/doi/pdf/10.1002/9783527677375.ch8}
  {https://onlinelibrary.wiley.com/doi/pdf/10.1002/9783527677375.ch8}
  \BibitemShut {NoStop}%
\bibitem [{\citenamefont {Igochine}\ \emph {et~al.}(2014)\citenamefont
  {Igochine}, \citenamefont {Gude}, \citenamefont {G\"unter}, \citenamefont
  {Lackner}, \citenamefont {Yu}, \citenamefont {{Barrera Orte}}, \citenamefont
  {Bogomolov}, \citenamefont {Classen}, \citenamefont {McDermott},
  \citenamefont {Luhmann~Jr.},\ and\ \citenamefont {{ASDEX Upgrade
  Team}}}]{IgochinePoP14}%
  \BibitemOpen
  \bibfield  {author} {\bibinfo {author} {\bibfnamefont {V.}~\bibnamefont
  {Igochine}}, \bibinfo {author} {\bibfnamefont {A.}~\bibnamefont {Gude}},
  \bibinfo {author} {\bibfnamefont {S.}~\bibnamefont {G\"unter}}, \bibinfo
  {author} {\bibfnamefont {K.}~\bibnamefont {Lackner}}, \bibinfo {author}
  {\bibfnamefont {Q.}~\bibnamefont {Yu}}, \bibinfo {author} {\bibfnamefont
  {L.}~\bibnamefont {{Barrera Orte}}}, \bibinfo {author} {\bibfnamefont
  {A.}~\bibnamefont {Bogomolov}}, \bibinfo {author} {\bibfnamefont
  {I.}~\bibnamefont {Classen}}, \bibinfo {author} {\bibfnamefont {R.~M.}\
  \bibnamefont {McDermott}}, \bibinfo {author} {\bibfnamefont {N.~C.}\
  \bibnamefont {Luhmann~Jr.}},\ and\ \bibinfo {author} {\bibnamefont {{ASDEX
  Upgrade Team}}},\ }\bibfield  {title} {\enquote {\bibinfo {title} {Conversion
  of the dominantly ideal perturbations into a tearing mode after a sawtooth
  crash},}\ }\href@noop {} {\bibfield  {journal} {\bibinfo  {journal} {Physics
  of Plasmas}\ }\textbf {\bibinfo {volume} {21}},\ \bibinfo {pages} {110702}
  (\bibinfo {year} {2014})}\BibitemShut {NoStop}%
\bibitem [{\citenamefont {Willensdorfer}\ \emph {et~al.}(2024)\citenamefont
  {Willensdorfer}, \citenamefont {Mitterauer}, \citenamefont {Hoelzl},
  \citenamefont {Suttrop}, \citenamefont {Cianciosa}, \citenamefont {Dunne},
  \citenamefont {Fischer}, \citenamefont {Leuthold}, \citenamefont {Puchmayr},
  \citenamefont {Samoylov}, \citenamefont {{Su\'arez L\'opez}}, \citenamefont
  {Wendler},\ and\ \citenamefont {{ASDEX Upgrade
  Team}}}]{WillensdorferNatPhys24}%
  \BibitemOpen
  \bibfield  {author} {\bibinfo {author} {\bibfnamefont {M.}~\bibnamefont
  {Willensdorfer}}, \bibinfo {author} {\bibfnamefont {V.}~\bibnamefont
  {Mitterauer}}, \bibinfo {author} {\bibfnamefont {M.}~\bibnamefont {Hoelzl}},
  \bibinfo {author} {\bibfnamefont {W.}~\bibnamefont {Suttrop}}, \bibinfo
  {author} {\bibfnamefont {M.}~\bibnamefont {Cianciosa}}, \bibinfo {author}
  {\bibfnamefont {M.}~\bibnamefont {Dunne}}, \bibinfo {author} {\bibfnamefont
  {R.}~\bibnamefont {Fischer}}, \bibinfo {author} {\bibfnamefont
  {N.}~\bibnamefont {Leuthold}}, \bibinfo {author} {\bibfnamefont
  {J.}~\bibnamefont {Puchmayr}}, \bibinfo {author} {\bibfnamefont
  {O.}~\bibnamefont {Samoylov}}, \bibinfo {author} {\bibfnamefont
  {G.}~\bibnamefont {{Su\'arez L\'opez}}}, \bibinfo {author} {\bibfnamefont
  {D.}~\bibnamefont {Wendler}},\ and\ \bibinfo {author} {\bibnamefont {{ASDEX
  Upgrade Team}}},\ }\bibfield  {title} {\enquote {\bibinfo {title}
  {{Observation of magnetic islands in tokamak plasmas during the suppression
  of edge-localized modes}},}\ }\href
  {https://doi.org/10.1038/s41567-024-02666-y} {\bibfield  {journal} {\bibinfo
  {journal} {Nature Physics}\ }\textbf {\bibinfo {volume} {24}},\ \bibinfo
  {pages} {1980--1988} (\bibinfo {year} {2024})}\BibitemShut {NoStop}%
\bibitem [{\citenamefont {Muraglia}\ \emph {et~al.}(2025)\citenamefont
  {Muraglia}, \citenamefont {Agullo}, \citenamefont {Dubuit}, \citenamefont
  {Bigu\'e},\ and\ \citenamefont {Garbet}}]{MuragliaJPP25}%
  \BibitemOpen
  \bibfield  {author} {\bibinfo {author} {\bibfnamefont {M.}~\bibnamefont
  {Muraglia}}, \bibinfo {author} {\bibfnamefont {O.}~\bibnamefont {Agullo}},
  \bibinfo {author} {\bibfnamefont {N.}~\bibnamefont {Dubuit}}, \bibinfo
  {author} {\bibfnamefont {R.}~\bibnamefont {Bigu\'e}},\ and\ \bibinfo {author}
  {\bibfnamefont {X.}~\bibnamefont {Garbet}},\ }\bibfield  {title} {\enquote
  {\bibinfo {title} {Multi-scale physics of magnetic reconnection in hot
  magnetized plasmas},}\ }\href@noop {} {\bibfield  {journal} {\bibinfo
  {journal} {Journal of Plasmas Physics}\ }\textbf {\bibinfo {volume} {91}},\
  \bibinfo {pages} {E19} (\bibinfo {year} {2025})}\BibitemShut {NoStop}%
\bibitem [{\citenamefont {{Villa}}\ \emph {et~al.}(2025)\citenamefont
  {{Villa}}, \citenamefont {{Dubuit}}, \citenamefont {{Agullo}},\ and\
  \citenamefont {{Garbet}}}]{VillaPoP25}%
  \BibitemOpen
  \bibfield  {author} {\bibinfo {author} {\bibfnamefont {D.}~\bibnamefont
  {{Villa}}}, \bibinfo {author} {\bibfnamefont {N.}~\bibnamefont {{Dubuit}}},
  \bibinfo {author} {\bibfnamefont {O.}~\bibnamefont {{Agullo}}},\ and\
  \bibinfo {author} {\bibfnamefont {X.}~\bibnamefont {{Garbet}}},\ }\bibfield
  {title} {\enquote {\bibinfo {title} {{Zonal fields as catalysts and
  inhibitors of turbulence-driven magnetic islands}},}\ }\href
  {https://doi.org/10.1063/5.0243358} {\bibfield  {journal} {\bibinfo
  {journal} {Physics of Plasmas}\ }\textbf {\bibinfo {volume} {32}},\ \bibinfo
  {eid} {010701} (\bibinfo {year} {2025})}\BibitemShut {NoStop}%
\bibitem [{\citenamefont {Smolyakov}\ \emph {et~al.}(1995)\citenamefont
  {Smolyakov}, \citenamefont {Hirose}, \citenamefont {Lazzaro}, \citenamefont
  {Re},\ and\ \citenamefont {Callen}}]{SmolyakovPoP95}%
  \BibitemOpen
  \bibfield  {author} {\bibinfo {author} {\bibfnamefont {A.~I.}\ \bibnamefont
  {Smolyakov}}, \bibinfo {author} {\bibfnamefont {A.}~\bibnamefont {Hirose}},
  \bibinfo {author} {\bibfnamefont {E.}~\bibnamefont {Lazzaro}}, \bibinfo
  {author} {\bibfnamefont {G.~B.}\ \bibnamefont {Re}},\ and\ \bibinfo {author}
  {\bibfnamefont {J.~D.}\ \bibnamefont {Callen}},\ }\bibfield  {title}
  {\enquote {\bibinfo {title} {{Rotating nonlinear magnetic islands in tokamak
  plasmas}},}\ }\href@noop {} {\bibfield  {journal} {\bibinfo  {journal}
  {Physics of Plasmas}\ }\textbf {\bibinfo {volume} {2}},\ \bibinfo {pages}
  {1581--1598} (\bibinfo {year} {1995})}\BibitemShut {NoStop}%
\bibitem [{\citenamefont {Wilson}\ \emph {et~al.}(1996)\citenamefont {Wilson},
  \citenamefont {Connor}, \citenamefont {Hastie},\ and\ \citenamefont
  {Hegna}}]{Wilson1996a}%
  \BibitemOpen
  \bibfield  {author} {\bibinfo {author} {\bibfnamefont {H.~R.}\ \bibnamefont
  {Wilson}}, \bibinfo {author} {\bibfnamefont {J.~W.}\ \bibnamefont {Connor}},
  \bibinfo {author} {\bibfnamefont {R.~J.}\ \bibnamefont {Hastie}},\ and\
  \bibinfo {author} {\bibfnamefont {C.~C.}\ \bibnamefont {Hegna}},\ }\bibfield
  {title} {\enquote {\bibinfo {title} {{Threshold for neoclassical magnetic
  islands in a low collision frequency tokamak}},}\ }\href
  {https://doi.org/10.1063/1.871830} {\bibfield  {journal} {\bibinfo  {journal}
  {Physics of Plasmas}\ }\textbf {\bibinfo {volume} {3}},\ \bibinfo {pages}
  {248--265} (\bibinfo {year} {1996})}\BibitemShut {NoStop}%
\bibitem [{\citenamefont {{Siccinio}}\ \emph {et~al.}(2011)\citenamefont
  {{Siccinio}}, \citenamefont {{Poli}}, \citenamefont {{Casson}}, \citenamefont
  {{Hornsby}},\ and\ \citenamefont {{Peeters}}}]{SiccinioPoP11}%
  \BibitemOpen
  \bibfield  {author} {\bibinfo {author} {\bibfnamefont {M.}~\bibnamefont
  {{Siccinio}}}, \bibinfo {author} {\bibfnamefont {E.}~\bibnamefont {{Poli}}},
  \bibinfo {author} {\bibfnamefont {F.~J.}\ \bibnamefont {{Casson}}}, \bibinfo
  {author} {\bibfnamefont {W.~A.}\ \bibnamefont {{Hornsby}}},\ and\ \bibinfo
  {author} {\bibfnamefont {A.~G.}\ \bibnamefont {{Peeters}}},\ }\bibfield
  {title} {\enquote {\bibinfo {title} {{Gyrokinetic determination of the
  electrostatic potential of rotating magnetic islands in tokamaks}},}\ }\href
  {https://doi.org/10.1063/1.3671964} {\bibfield  {journal} {\bibinfo
  {journal} {Physics of Plasmas}\ }\textbf {\bibinfo {volume} {18}},\ \bibinfo
  {eid} {122506} (\bibinfo {year} {2011})}\BibitemShut {NoStop}%
\bibitem [{NoteI()}]{NoteI}%
  \BibitemOpen
  \bibinfo {note} {For the simulation results reported in figure~\ref
  {fig:MTMTeProfs}), poloidally-resolved profiles are unfortunately not
  available; flux-surface-averaged profiles are shown instead.}\BibitemShut
  {Stop}%
\end{thebibliography}%
\end{document}